\documentclass[12pt]{iopart}

\usepackage{iopams}  
\usepackage{graphicx}
\newcommand\diff[1]{{\rm d}#1}
\newcommand\ddiff[1]{{\rm d^2}#1}
\newcommand\dddiff[1]{{\rm d^3}#1}
\newcommand\ddimdiff[1]{{\rm d}^D#1}
\newcommand\dddimdiff[1]{{\rm d}^{D-1}#1}
\newcommand\wrmi[2]{\ensuremath{{#1}_{\rm #2}}}
\newcommand\setR{\mathbb{R}}

\newcommand\operatorname[1]{{\mathrm{#1}}}
\newcommand\eqref[1]{(\ref{#1})}
\newcommand\corr{\operatorname{corr}}
\newcommand\cov{\operatorname{cov}}
\newcommand\bodyI{K^{(i)}}
\newcommand\Voronoi{Voronoi}

\begin{document}
%\linenumbers
\title[Local Anisotropy of Fluids]{Local Anisotropy of Fluids using Minkowski Tensors}

\author{S.\,C.~Kapfer$^{1,*}$, W.~Mickel$^1$, F.\,M.~Schaller$^1$, M.~Spanner$^1$, C.~Goll$^1$,
    T.~Nogawa$^2$, N.~Ito$^2$, K.~Mecke$^1$, and G.\,E.~Schr\"oder-Turk$^1$}

\address{$^1$Institut f\"ur Theoretische Physik,
    Friedrich-Alexander-Universit\"at Erlangen-N\"urnberg,
    Staudtstr.~7, D-91058 Erlangen, Germany}
\address{$^2$Department of Applied Physics,
    School of Engineering, The University of Tokyo, Japan}
\ead{$^*$Sebastian.Kapfer@physik.uni-erlangen.de}

%\FIXME{TITEL: Gerd: "Anisotropy of liquid free volume cells"}
\begin{abstract}
Statistics of the free volume available to individual particles have previously been studied
for simple and complex fluids, granular matter, amorphous solids, and structural glasses.
Minkowski tensors provide a set of shape measures that are based
on strong mathematical theorems and easily computed for polygonal and
polyhedral bodies such as free volume cells (Voronoi cells).
They characterize the local structure beyond the two-point
correlation function and are suitable to define indices $0\leq \beta_\nu^{a,b}\leq 1$ of local anisotropy.
Here, we analyze the statistics of Minkowski tensors for configurations of
simple liquid models, including the ideal gas (Poisson point process), the
hard disks and hard spheres ensemble, and the Lennard-Jones fluid.
We show that Minkowski tensors
provide a robust characterization of local anisotropy, which ranges from
$\beta_\nu^{a,b}\approx 0.3$ for vapor phases to $\beta_\nu^{a,b}\rightarrow
1$ for ordered solids.
We find that for fluids, local anisotropy decreases monotonously with
increasing free volume  and randomness of particle positions.
Furthermore, the local anisotropy indices $\beta_\nu^{a,b}$ are sensitive to
structural transitions in these simple fluids, as has been 
previously shown in granular systems for the transition from
loose to jammed bead packs.
\end{abstract}

%Uncomment for PACS numbers title message
%\pacs{00.00, 20.00, 42.10}
% Keywords required only for MST, PB, PMB, PM, JOA, JOB? 
%\vspace{2pc}
%\noindent{\it Keywords}: Article preparation, IOP journals
% Uncomment for Submitted to journal title message
%\submitto{\JPA}
% Comment out if separate title page not required
\maketitle
%
%\FIXME{We're going to submit to Journal of Statistical Mechanics: Theory and Experiment}
%\FIXME{can upload a copy on arXiv}
%
%
\section{Introduction}
As new experimental techniques become available, more detailed information on the
nature of the fluid state can be accessed.  New scattering experiments
reveal the local ordering of fluids \cite{PeterWochner07142009,ReichertFiveFoldLead},
and confocal microscopy allows for individual particles to be tracked \cite{EricRWeeks01282000}.
This richness of new data calls for new methods to characterize the morphology
of equilibrium and nonequilibrium fluids and related systems.

\begin{figure}[b]
\centering
\begin{tabular}{llll}
\includegraphics[clip,width=.197\linewidth]{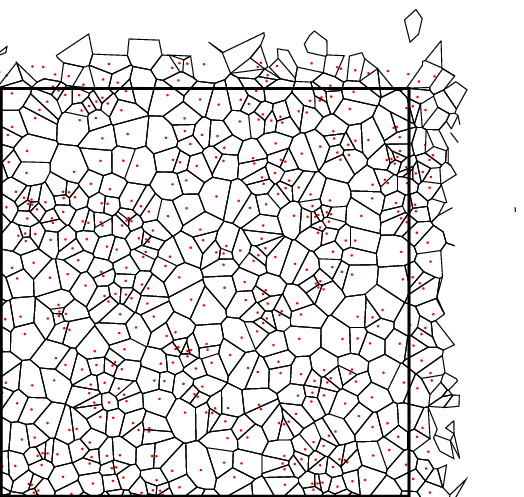}&
\includegraphics[clip,width=.197\linewidth]{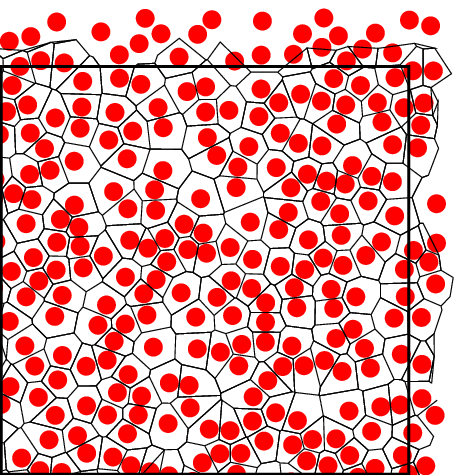}&
\includegraphics[clip,width=.197\linewidth]{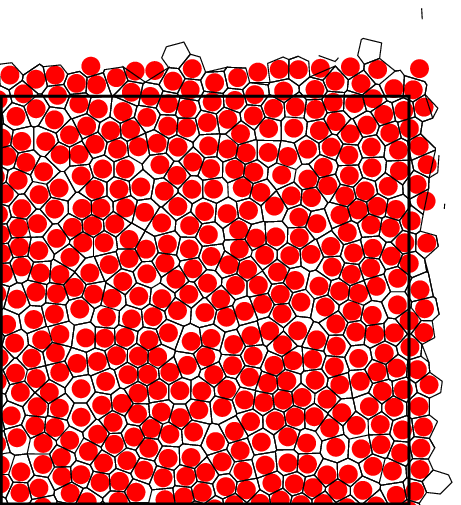}&
\includegraphics[clip,width=.197\linewidth]{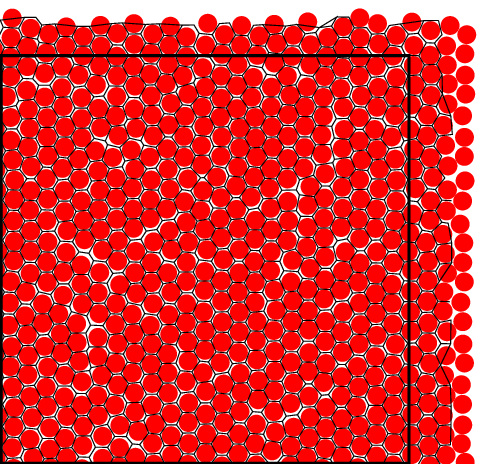}\\
\includegraphics[trim=1.32cm 0 .3cm 0,width=.2\linewidth]{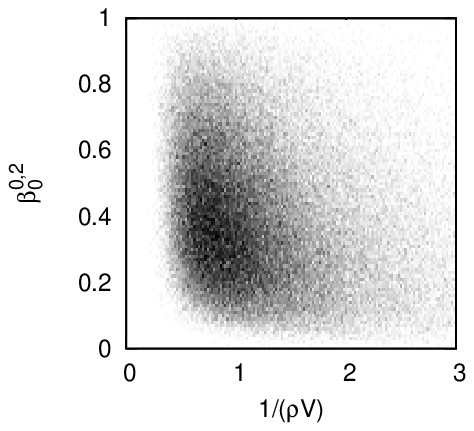}&
\includegraphics[clip,trim=1.32cm 0 .3cm 0,width=.2\linewidth]{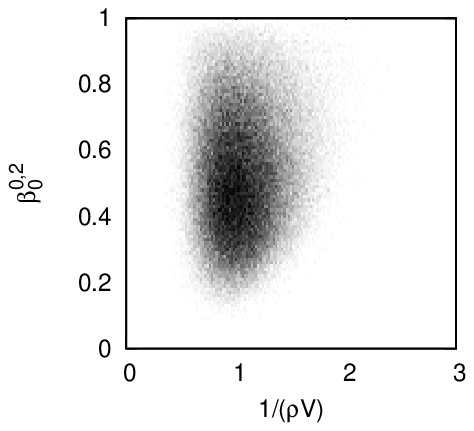}&
\includegraphics[clip,trim=1.32cm 0 .3cm 0,width=.2\linewidth]{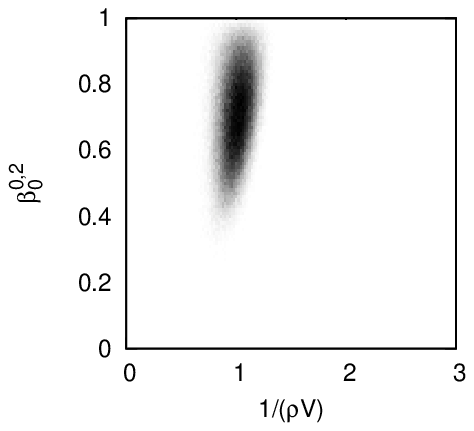}&
\includegraphics[clip,trim=1.32cm 0 .3cm 0,width=.2\linewidth]{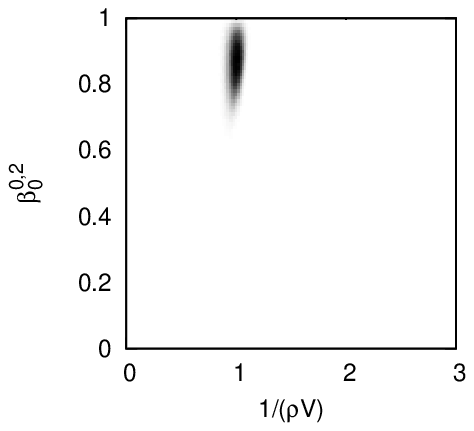}
\end{tabular}

\caption{Histograms of local density ($x$ axis) and Voronoi cell anisotropy index ($y$ axis) in
a Poisson point pattern, a dilute hard disks fluid ($\eta=32\%$), and hard disks
configurations below and above the phase transition ($\eta = 61\%$, $76\%$).  The shape of
the distribution changes characteristically with the packing density.
}
\label{fig:first-page-picture}
\end{figure}

While fluids of spherical particles usually are globally isotropic,
the local environments of the particles may be anisotropic (fig.~\ref{fig:first-page-picture}).
Local anisotropy in jammed bead packs has recently been studied in the framework
of Minkowski tensors \cite{NonScience:2010}. 
%It is well-known that ellipsoidal
%particles pack denser than spherical ones \cite{DonevCisseSachsVarianoStillingerConellyTorquatoChaikin:2004},
%and it is therefore natural to assume a degree of local anisotropy of the void space of a bead pack.
Local anisotropy may, in fact, play a crucial role both for dynamical and equilibrium
properties of particle systems.  At the very least, measures of local
anisotropy can be used as a local order parameter to discriminate between the
various equilibrium and nonequilibrium states of an assembly of particles.
This article extends the analysis given in \cite{NonScience:2010} to important
model fluids, namely the ideal gas (Poisson point process), the hard spheres and
hard disks ensembles, and Lennard-Jones fluids.  We demonstrate that the local anisotropy
yields distinct evidence of the phase behavior of simple fluids; furthermore, we
show that Minkowski tensors provide a robust means to its characterization.

As a measure of local geometry in particulate matter, local anisotropy evidently competes
with existing measures based on nearest-neighbor analysis, including
the bond-orientational order parameters
\cite{PhysRevE.55.6855,TorquatoTruskettDebenedetti:2000,PhysRevLett.99.215701,LechnerDellago:2008},
the isoperimetric ratio \cite{PhysRevLett.94.040601,PhysRevLett.96.258001}
and related quantities \cite{StarrSastryDouglasGlotzer:2002},
and tensorial measures such as Edwards' configurational tensor \cite{Edwards2001162},
quadrons \cite{PhysRevLett.90.114303},
and fabric / texture tensors \cite{DurandIceCore:2004}.
These measures are, while undoubtedly widely and successfully applied,
reliant on identifying suitable neighborhoods.  Neighborhoods, by their
very nature, change discontinuously, in contrast to the shape measures of Voronoi cells
used in the present study.  The volume distributions of \Voronoi{} cells have been
studied \cite{Meijering:1953,Bernal:Nature1957,OkabeVoronoiBook:2000}
and the shape of \Voronoi{} cells is their next important property.
Scalar measures of shape such as the isoperimetric ratio
contain some signature of anisotropy;  however, they do not discern between the generally distinct
but often correlated concepts of asphericity (meaning deviation form a spherical shape)
and anisotropy (meaning elongation or directedness).
It is our hypothesis that explicit anisotropy measures such as the Minkowski tensors
will turn out similarly useful for the identification of phase transitions in fluids and
other particulate systems, while being more generic and robust in their definition
than measures based on neighborhoods.  Explicit anisotropy measures may also be
more easily generalized to aspherical particles, relevant to liquid crystalline phases
\cite{PhysRevLett.52.287,PhysRevA.41.3237,EllisonMichelBarmesCleaver:2006}.

The article is organized as follows:
In section \ref{sec:minktensors}, we restate briefly the definition of the
Minkowski tensors. Section \ref{sec:aniso-chara} discusses how Minkowski tensors may
be used to characterize anisotropy. In section \ref{sec:voronoianiso}, we
use Minkowski tensors of \Voronoi{} tessellations to characterize the local
anisotropy in a point pattern, such as the configuration of particles in a fluid.
Sections \ref{sec:ppp} and \ref{sec:disks-and-spheres} gives the results of the anisotropy analysis for
an ideal gas, described by the Poisson point process, and equilibrium systems of hard spheres and disks.
Section \ref{sec:lennard-jones} applies the same analysis
to point particles interacting with a Lennard-Jones potential.
Section \ref{sec:misc} shows that Minkowski tensors are both robust and sensitive
measures of local geometry by applying them to the Einstein solid model.

\section{Anisotropy Characterization by Minkowski Tensors}
\subsection{Minkowski Tensors}
\label{sec:minktensors}
The description of shape requires the choice of adequate quantities
that may be associated with the shape in question.
Minkowski functionals, also known as \emph{Querma\ss{}integrale} or
\emph{intrinsic volumes}, are scalar measures of shape defined
in convex and integral geometry, based on support measures \cite{Hadwiger:1957,Santalo:1976}.
Equivalently, they may be introduced as curvature-weighted integrals,
as in this section.

Given a compact subset of $D$-dimensional Euclidean space $K\subset \setR^D$,
called a \emph{body}, with a sufficiently smooth boundary,
the Minkowski functionals in two and three dimensions are defined as
\begin{eqnarray}
    W_0(K)&:=& \int_K \ddimdiff{r} \\
 W_\nu(K) &:=& {\frac 1D}\int_{\partial K}
               \dddimdiff{r} \; G_\nu(\vec r), \quad 1\leq \nu \leq D,
\label{def:minkowskis}
\end{eqnarray}
where $G_\nu(\vec r)$ is a symmetric combination of the principal curvatures
$\kappa_i$ of 
the boundary $\partial K$.  In two dimensions, $G_1=1, G_2=\kappa$,
(there is only one curvature)
and in three dimensions $G_1=1, G_2 = \frac 12 (\kappa_1+\kappa_2), G_3 = \kappa_1\kappa_2$.
The Minkowski functionals are, up to conventional
prefactors, the area, the circumference, and  the Euler number (in 2D) and
the volume, the surface area, the integral mean curvature, and the Euler number
(in 3D) of the body $K$.
The set of Minkowski functionals has a number of properties which
make them useful for shape analysis of physical systems \cite{Mecke:2000}.
In particular, by Hadwiger's characterization theorem \cite{Hadwiger:1957},
the set of Minkowski functionals
spans the space of additive, conditionally continuous and motion invariant
functionals of $K$.
For such functionals, the Minkowski functionals contain all the relevant
morphological information about the shape of $K$.
%In this sense, the Minkowski functionals contain the complete
%motion-invariant information about the shape of $K$.

A natural generalization of Minkowski functionals are Minkowski tensors,
which arise by including tensor products of the position vector $\vec r$
and the outer boundary normal $\vec n$ in the integrals in eq.~\eqref{def:minkowskis}.
In analogy to the scalar case, the Minkowski tensors
span the space of additive, conditionally continuous
and motion-\emph{covariant} functionals \cite{Alesker:1999}.
Knowledge of all the Minkowski tensors provides a more complete morphological
description of the body $K$ than the scalar properties alone, and allows
for tensor-valued functionals of $K$ to be computed.
The set of Minkowski tensors contains, analogous to the scalar case,
complete morphological information about the shape of $K$.
A basis of the space of tensor-valued functionals of rank two is given in
table \ref{tab:minktens}.  In addition to the scalar Minkowski functionals,
four Minkowski tensors (six in 3D) are required to span the space.
\begin{table}
\centering
\begin{tabular}[t]{ll|l}
& Tensor & Matrix element $i, j$ \\\hline
$D=2$ &
    $E \cdot W_0$ &$\delta_{ij}\cdot\int_K \ddiff r $ \\
&
    $E \cdot W_1$ &$\frac 12 \delta_{ij}\cdot\int_{\partial K} \diff r $ \\
&
    $E \cdot W_2$ &$\frac 12 \delta_{ij}\cdot\int_{\partial K} \diff r \;\kappa $ \\
&
    \\\hline
&
    $W_0^{2,0}$ &$\int_K \ddiff r \; r_i r_j $ \\
&
    $W_1^{2,0}$ &$\frac 12 \int_{\partial K} \diff r \; r_i r_j $ \\
&
    $W_2^{2,0}$ &$\frac 12 \int_{\partial K} \diff r \; \kappa \; r_i r_j $ \\
&
    \\\hline
&
    $W_1^{0,2}$ &$\frac 12 \int_{\partial K} \diff r \; n_i n_j $ \\
&
\end{tabular}\;\;%
\begin{tabular}[t]{ll|l}
& Tensor & Matrix element $i, j$ \\\hline
$D=3$ &
    $E \cdot W_0$ &$\delta_{ij}\cdot\int_K \dddiff r $ \\
&
    $E \cdot W_1$ &$\frac 13\delta_{ij}\cdot\int_{\partial K} \ddiff r $ \\
&
    $E \cdot W_2$ &$\frac 16\delta_{ij}\cdot\int_{\partial K} \ddiff r \;\bigl(\kappa_1+\kappa_2\bigr) $ \\
&
    $E \cdot W_3$ &$\frac 13\delta_{ij}\cdot\int_{\partial K} \ddiff r \;\kappa_1\kappa_2 $ \\\hline
&
    $W_0^{2,0}$ &$\int_K \dddiff r \; r_i r_j $ \\
&
    $W_1^{2,0}$ &$\frac 13\int_{\partial K} \ddiff r \; r_i r_j $ \\
&
    $W_2^{2,0}$ &$\frac 16\int_{\partial K} \ddiff r \; \bigl(\kappa_1+\kappa_2\bigr) \; r_i r_j $ \\
&
    $W_3^{2,0}$ &$\frac 13\int_{\partial K} \ddiff r \;\kappa_1\kappa_2 \; r_i r_j $ \\\hline
&
    $W_1^{0,2}$ &$\frac 13\int_{\partial K} \ddiff r \; n_i n_j $\\
&
    $W_2^{0,2}$ &$\frac 16\int_{\partial K} \ddiff r \; \bigl(\kappa_1+\kappa_2\bigr) \; n_i n_j $
\end{tabular}

\caption{A basis for the space of rank-two additive, conditionally continuous,
motion-covariant functionals in terms of Minkowski functionals $W_\nu$ and Minkowski tensors
$W_\nu^{a,b}$.
$E$ is the unit matrix,
$\kappa=\kappa(\vec r)$ the local curvature of the boundary contour $\partial K$
at point $\vec r$,
$\kappa_1(\vec r)$ and $\kappa_2(\vec r)$
the principal curvatures of the boundary surface $\partial K$,
$\vec n(\vec r)$ the outer boundary normal at point $\vec r$.}
\label{tab:minktens}
\end{table}

For the mathematical details we refer the reader elsewhere
\cite{SchroederTurkKapferBreidenbachBeisbartMecke:2009,MickelBreidenbachMeckeSchroederTurk:2008,HugSchneiderSchuster:2008},
and instead only give a single example, namely the Minkowski tensor
$W_1^{0,2}$ of a planar body $K$ which is given by
\begin{equation}
    W_1^{0,2}(K) = {\textstyle\frac 12}\int_{\partial K} \diff{r}
    \left(\begin{array}{ccc} n_1 n_1 & n_1 n_2 \\ 
                             n_2 n_1 & n_2 n_2 \end{array}\right)
\end{equation}
where $n_i$ are the components of the outer normal vector
at the boundary $\partial K$ of $K$, and $\diff{r}$ is the infinitesimal
line element on the boundary $\partial K$.
$W_1^{0,2}(K)$ can now be written in terms of three independent scalar boundary integrals
\begin{equation}
    W_1^{0,2}(K) = 
    \left(\begin{array}{ccc} I_{11}(K) & I_{12}(K) \\ 
                             I_{12}(K) & I_{22}(K) \end{array}\right).
\label{eq:explicit-matrix-form}
\end{equation}
Importantly for applications, the Minkowski tensors of a triangulated
body (in the planar case, a polygon) can be written as a simple sum over all
facets and thus calculated in $\mathcal O(N)$ time.  For a polygon, the $I_{ij}(K)$
in eq.~\eqref{eq:explicit-matrix-form} are given by the three sums
\begin{equation}
    I_{ij}(K) = {\textstyle\frac 12} \sum_f L^{(f)} n^{(f)}_i n^{(f)}_j
\label{eq:w102sumrepr}
\end{equation}
where the index $f$ labels the edges of the polygon, and $L^{(f)}$, $\vec n^{(f)}$
are their lengths and outer normals respectively.  An analogous formula holds in three
dimensions, replacing edge lengths by facet areas.
Explicit formulae for the fast and simple computation of the complete set of
Minkowski tensors ($W_0^{2,0}$, $W_1^{2,0}$, $W_2^{2,0}$, $W_1^{0,2}$ in 2D and 
$W_0^{2,0}$, $W_1^{2,0}$, $W_2^{2,0}$, $W_3^{2,0}$, $W_1^{0,2}$, $W_2^{0,2}$
in 3D) can be found in
refs.~\cite{SchroederTurkKapferBreidenbachBeisbartMecke:2009,MickelBreidenbachMeckeSchroederTurk:2008}.
%
%Explicit formulae for the fast and simple computation of all other rank-two
%Minkowski tensors of a polytope can be found in
%refs.~

Similar to the tensor of inertia, some of the Minkowski tensors, $W_\nu^{2,0}$,
do not depend only on the shape of $K$ but also on its
location.  A suitable origin has to be chosen for computation of the Minkowski
tensors for a given body.  The tensors including only normal vectors,
$W_\nu^{0,2}$, are translation invariant.

\subsection{Anisotropy Characterization of Shapes}
\label{sec:aniso-chara}%
Having assigned a number of tensorial shape descriptors to each body $K$,
we can now define an anisotropy index for each body $K$
and each Minkowski tensor $W_\nu^{a,b}(K)$
by the dimensionless ratio
\begin{equation}
    \beta_\nu^{a,b}(K) := \frac{\wrmi\lambda{min}(W_\nu^{a,b}(K))}{\wrmi\lambda{max}(W_\nu^{a,b}(K))},
\label{def:beta}
\end{equation}
where $\wrmi\lambda{min}(W)$, $\wrmi\lambda{max}(W)$ are the smallest and largest
eigenvalues of the $D\times D$ matrix $W$; $D$ the spatial dimension.
An anisotropy index of $\beta=1$ indicates an \emph{isotropic body}, and values ranging from
1 to 0 correspond to increasing degrees of anisotropy.  
The dimensionless anisotropy index is a pure shape measure; it is invariant
under isotropic scaling of $K$, i.\,e.~$\beta_\nu^{a,b}(s K) = \beta_\nu^{a,b}(K)$,
for all $s>0$.
    
Note that, in the framework of this analysis, we consider any body isotropic
that has Minkowski tensors proportional to the unit matrix.
Bodies with $n$-fold rotation symmetries through the origin ($n\geq 3$) have degenerate eigenvalues
due to the additivity of the Minkowski tensors; as a consequence, many well-known highly
symmetric bodies have $\beta=1$.  This includes, in two dimensions, the sphere and square,
and any regular polygon; in three dimensions, sphere and cube, and the fcc unit cell
(truncated octahedron) centered at the origin are examples of isotropic bodies.

\subsection{Characterization of Point Patterns}
\label{sec:voronoianiso}%
We characterize local anisotropy of point configurations, representing for example
the centroids of fluid particles, by computing the anisotropy indices $\beta$
of their Voronoi cells.
Each point $g^{(i)}$ in the configuration,
termed \emph{germ}, is assigned a \Voronoi{} cell $\bodyI{}$ that consists of
all points in $\setR^D$ closer to $g^{(i)}$ than to any other germ $g^{(j)}\not=g^{(i)}$.
From any point configuration $\{g^{(i)}\}$, $i = 1$, \dots, $N$ we thus obtain
a set of $N$ bodies $\{\bodyI{}\}$,
and consequently, $N$ Minkowski tensors $\{W_\nu^{a,b}(\bodyI{})\}$.
For the translation invariant Minkowski tensors, the choice of origin is irrelevant;
for the others, the germ point $g^{(i)}$ is chosen as the origin
to compute the tensor $W_\nu^{a,b}(\bodyI{})$.
\Voronoi{} cells are frequently used to define the \emph{free volume} available
to an individual particle in a fluid configuration, and in glassy and supercooled states,
for example \cite{StarrSastryDouglasGlotzer:2002}.
The same concept has been used in granular matter for static packings \cite{AsteDiMatteo:2008}.
The \Voronoi{} tessellations of a set of points is computed using readily
available software \cite{BarberDobkinHuhdanpaa:1996}.  In contrast to the neighborhood topology, the
shapes of the \Voronoi{} cells are a continuous function of the point pattern.

For the remainder of the paper, the index $i$ labels the germ point in the point
configuration.
Since we study point configurations consisting of a large number of points,
all results will be of statistical nature.
Averages over the germs in one or more configurations are denoted by $\mu({}\cdot{}{})$;
for example, $\mu(\beta_\nu^{a,b}(\bodyI{}))
= \frac 1N  \cdot \sum_i \beta_\nu^{a,b}(\bodyI{})$ is the average anisotropy index of \Voronoi{}
cells, with the average taken over all the germs. In addition,
$\sigma({}\cdot{}{})$ denotes the standard deviation;
for the observable $X^{(i)}$, it is computed from
$\sigma^2(X^{(i)}) = (N-1)^{-1} \cdot \sum_i (X^{(i)} - \mu(X^{(i)}))^2 $.
For brevity, we identify $\mu(\beta_\nu^{a,b}):=\mu(\beta_\nu^{a,b}(\bodyI{}))$.

\section{Poisson Point Process (Ideal Gas)}\label{sec:ppp}%
The simplest model of a fluid is the ideal gas;  molecules in the gas do not
interact, and as a consequence, their locations $g^{(i)}$ are uncorrelated.
Mathematically, ideal gas configurations are modeled by a Poisson point process (PPP)
\cite{SchneiderWeil:2008}.  The only free parameter in the Poisson point process
is the \emph{intensity} $\rho$, which is equal to the expectation value of the number
of particles per unit volume.  This parameter can be removed by a rescaling
the coordinates in the system.  Since the anisotropy indices
$\beta_\nu^{a,b}$, as defined above, are invariant under scaling of the body,
the measured degree of anisotropy is independent of the intensity $\rho$.

\Voronoi{} tessellations are difficult to treat analytically. Thus, we
estimate the parameters of the $\beta_\nu^{a,b}$ distributions numerically.
For this purpose, ten realizations of Poisson process on a
square $[0;\,1)^2$ and on a cube $[0;\,1)^3$ are generated by first drawing
the number of germs from a Poisson
distribution with mean 16384 and subsequently placing the germs in the square
or cube; germ point coordinates are drawn independently from uniform distributions in
on $[0;\,1)$.  The \Voronoi{} tessellation is computed for each configuration with
periodic boundary conditions, and Minkowski tensors and functionals are computed
for the individual \Voronoi{} cells.

Table~\ref{tbl:scalars} shows results for the average scalar Minkowski
functionals of Poisson-\Voronoi{} cells.  Analytical expressions are known for $W_0$ and $W_1$,
and we find numerical values close to the exact results \cite{Meijering:1953,OkabeVoronoiBook:2000}.
The authors are not aware of an analytical expression for the
average integral mean curvature $\mu(W_2)$ of Poisson-\Voronoi{} cells; the same
is true for the standard deviations of the respective distributions.  The
scalar Minkowski functionals (area, perimeter in 2D and volume, surface area,
integral mean curvature in 3D) are found to be compatible with generalized Gamma
distributions in agreement with the
literature (see fig.~\ref{fig:scalar-prop-distros} in appendix and refs.~\cite{Hanson:1983,OkabeVoronoiBook:2000}).

\begin{table}
\centering
\begin{tabular}{cc|lll}
      & Fct. & $\mu(W_\nu)$ & exact $\mu(W_\nu)$ & $\sigma(W_\nu)$\\[.3em]\hline
$D=2$ & $W_0$ & $1.00\rho^{-1}$ & $\rho^{-1}$ & $0.53 \rho^{-1}$ \\
      & $W_1$ & $2.00\rho^{-1/2}$ & $2\rho^{-1/2}$ & $0.49 \rho^{-1/2}$ \\[.3em]\hline
$D=3$ & $W_0$ & $1.00\rho^{-1}$ & $\rho^{-1}$ & $0.42 \rho^{-1}$\\
      & $W_1$ & $ 1.94\rho^{-2/3}$& $\frac 43\smash{\left(\frac{4\pi}3\right)}^{1/3}\;\Gamma\!\left(\frac 53\right) \rho^{-2/3}$ & $0.49\rho^{-2/3}$ \\
      & $W_2$ & $3.05\rho^{-1/3}$ & $-$  & $0.78\rho^{-1/3}$
\end{tabular}

\caption{Scalar Minkowski functionals of \Voronoi{} cells induced by a Poisson
point process. $\rho$ is the intensity of the point process, $\Gamma$ the Gamma function,
and $D$ the spatial dimension.
Exact results are taken from \cite{OkabeVoronoiBook:2000}, numerical values estimated 
in the present study from 160,000 random \Voronoi{} cells.}
\label{tbl:scalars}
\end{table}

\begin{figure}
\includegraphics{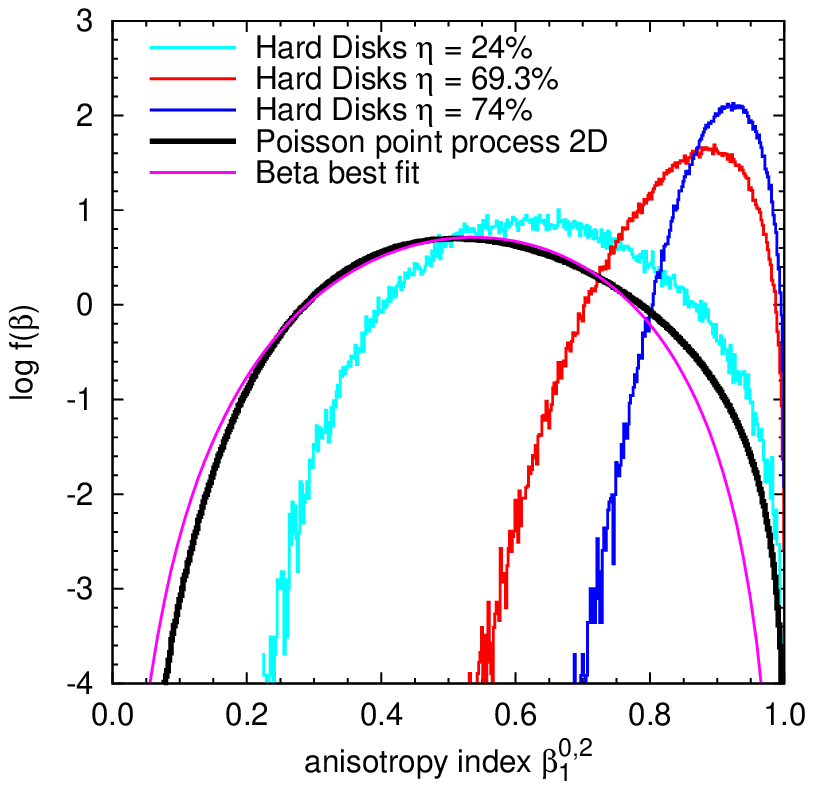}\includegraphics{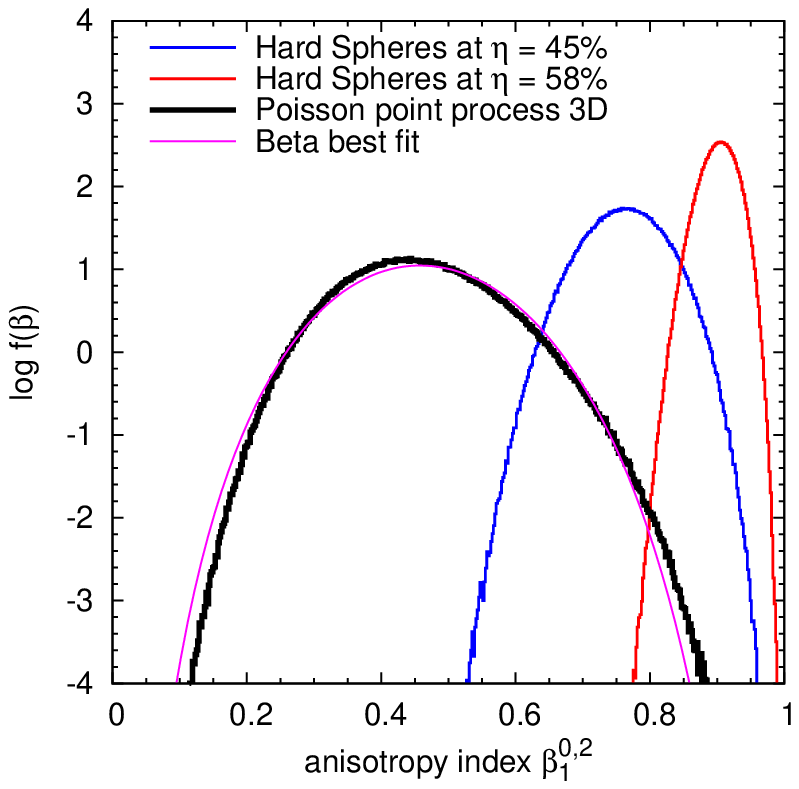}

\caption{Distribution of the anisotropy index $\beta_1^{0,2}$ in configurations
of hard disks (left) and spheres (right) of various packing fractions. The Poisson
limit $\eta\rightarrow 0$ is indicated by the bold black line.  Also shown is the
best fit of a Beta distribution $\propto x^m (1-x)^n$ to the Poisson limit.  }
%\FIXME{Parameter betaverteilungen}
%\FIXME{Linien in s/w besser unterscheidbar}
%
\label{fig:beta-distributions}
\label{fig:poisson-beta-beta-distr}
\end{figure}
The mean and standard deviation of the distribution of the anisotropy indices
$\beta_\nu^{a,b}$ of the Poisson-\Voronoi{} cells are shown in table~\ref{tbl:betapoisson}. 
Since the anisotropy indices are invariant under scaling, the length scale
$\rho$ does not enter, and the tabulated values are fingerprints of a point pattern
with uniformly and independently distributed germ points.
The full distribution of $\beta_1^{0,2}$ values is shown in fig.~\ref{fig:beta-distributions}
(bold lines).  In the 3D case, it is approximately a Beta distribution,  which
is remarkable because the Beta distribution is closely related to the
Gamma distribution\footnote{If the random variables $X$, $Y$ are independent Gamma variates with the same
length scale $\theta$ of the Gamma distribution, then $X/(X+Y)$ is a Beta variate.
In the case of $\beta_\nu^{a,b}$, $X$ and $Y$ correspond to the smallest eigenvalue $\wrmi\lambda{min}$
and the increment $\wrmi\lambda{max}-\wrmi\lambda{min}$, which are approximately Gamma distributed,
but not independent variables.}.  The agreement is less good in the 2D case.
%
%  There is a systematic effect in the scalars of the actual number of Poisson points.
%  For more precise results, this would have to be taken into account.
%
\begin{table}
\centering
\begin{tabular}[t]{cc|ll}
      & Tensor & $\mu(\beta_\nu^{a,b})$ & $\sigma(\beta_\nu^{a,b})$\\[.3em]\hline
$D=2$ & $W_0^{2,0}$ & 0.396  & 0.201 \\
      & $W_1^{2,0}$ & 0.463  & 0.197 \\
      & $W_2^{2,0}$ & 0.363  & 0.200 \\
      & & & \\\hline
      & $W_1^{0,2}$ & 0.540  & 0.180 \\
\end{tabular}
\hspace{.8cm}
\begin{tabular}[t]{cc|ll}
      & Tensor & $\mu(\beta_\nu^{a,b})$ & $\sigma(\beta_\nu^{a,b})$\\[.3em]\hline
$D=3$ & $W_0^{2,0}$ & 0.344  & 0.1330 \\
      & $W_1^{2,0}$ & 0.405  & 0.1330 \\
      & $W_2^{2,0}$ & 0.355  & 0.1301 \\
      & $W_3^{2,0}$ & 0.292  & 0.1261 \\\hline
      & $W_1^{0,2}$ & 0.457  & 0.1251 \\
      & $W_2^{0,2}$ & 0.694  & 0.0871
\end{tabular}

\caption{Mean and standard deviation of the anisotropy indices $\beta_\nu^{a,b}$
of the \Voronoi{} cells
induced by a Poisson point process.  The intensity $\rho$ of the
Poisson point process does not affect the dimensionless anisotropy indices.}
\label{tbl:betapoisson}
\end{table}
\section{Equilibrium Hard Spheres and Disks}
\begin{figure}
\centering
\includegraphics{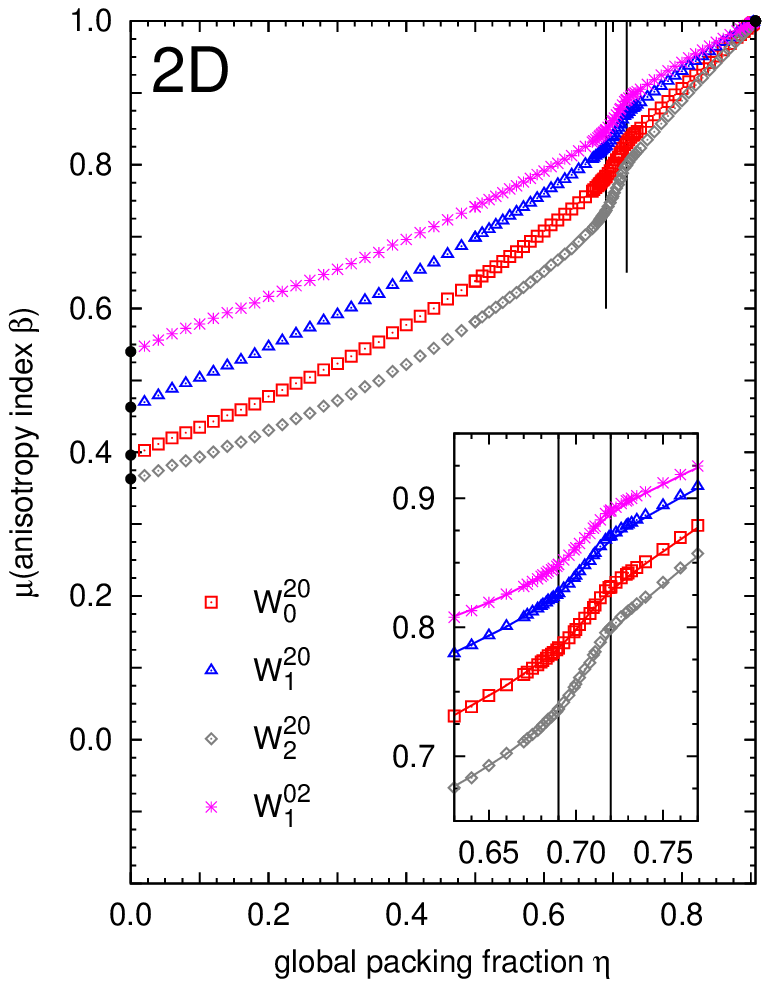}\includegraphics{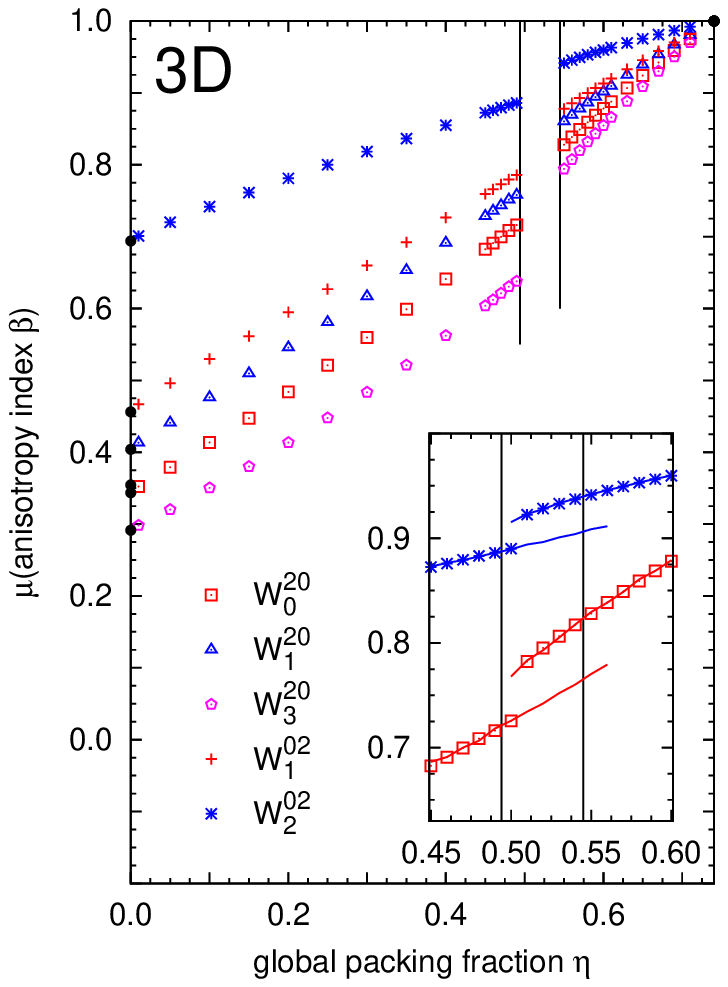}
\caption{Average anisotropy indices $\beta$ for equilibrium ensembles of hard disks
(left) and spheres (right) vs.~packing fraction $\eta$.
In the limit $\eta=0$, the hard spheres anisotropies are identical to those
of the Poisson process (solid bullets).
%Solid bullets on the plot frames represent the limits of the Poisson point process
%$\eta = 0$ and the triangular/fcc lattice $\eta=\wrmi{\eta}{max}$.

In 2D (left-hand side), changes of slope at the transition densities (magnified in inset)
are observed.  In 3D (right-hand side)
the average $\beta$ displays a discontinuity at the first-order phase transition.
The values of $\beta_2^{2,0}$ are very close to those of $\beta_0^{2,0}$ and have
been omitted for the sake of clarity.
The inset shows metastable states in or close to the coexistence region; symbols are
Monte Carlo results, lines molecular dynamics simulations.
}
\label{fig:hard-mu-of-beta}
\end{figure}

\begin{figure}
\centering
\includegraphics{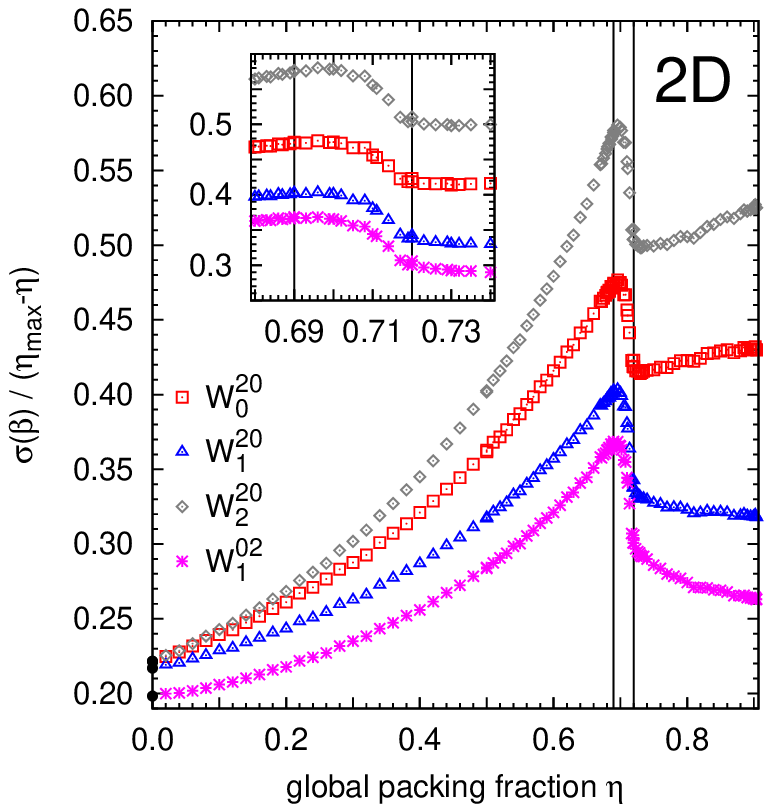}\includegraphics{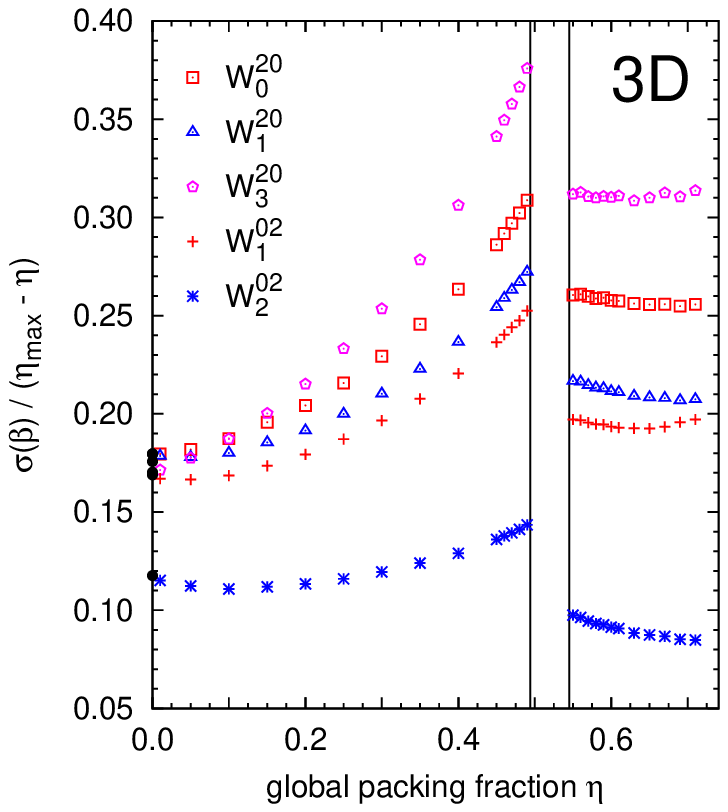}

\caption{Standard deviations of the $\beta$ distribution for equilibrium hard disks (left)
and spheres (right).  A linear decrease as $\eta\rightarrow\wrmi{\eta}{max}$ as been divided out.
All data points calculated from Monte Carlo simulations.
In the hard disks system, $\sigma(\beta)$ is continuous across the phase transition (magnified in inset).
The hard spheres system show a discontinuity.
The values of $\beta_2^{2,0}$ are very close to those of $\beta_0^{2,0}$ and have
been omitted for the sake of clarity.
Solid bullets on the plot frames represent the limit of the Poisson point process
$\eta = 0$.
}
\label{fig:hard-sigma-of-beta}
\end{figure}
\label{sec:disks-and-spheres}%
The hard disks and hard spheres model is beyond doubt one of the most important
generalizations of the ideal gas.  This model, consisting in spherical particles
without interaction other than elastic hard repulsion, possesses a fluid--solid
phase transition \cite{Alder:1957,Wood:1957}
and -- in the 3D case -- vitreous and jammed states \cite{PhysRevE.64.021506}.
In contrast to the ideal gas discussed above, the hard spheres model does
have an intrinsic length scale by virtue of the particle radius and packing fraction $\eta$.
This work focuses on monodisperse particles, even though the method is applicable to more
general models when one replaces the \Voronoi{} tessellation by a more general, for example
the Apollonius tessellation \cite{OkabeVoronoiBook:2000}.
%\FiXME{Cite Glotzer here?}

Two different procedures have been employed to generate realizations of
hard disks and hard spheres configurations:  First, Monte Carlo (MC) simulations
using alternating local Metropolis moves of single particles
and collective cluster moves, specifically 
swaps of clusters between the configuration and its inversion at a
randomly chosen point \cite{Krauth:2006,Dress:1995}.  Second, a 2D and a 3D event-driven
molecular dynamics (MD) code solving Newton's equations for a set of particles
interacting via a hard spheres potential \cite{Rapaport:2004,Masaharu-Isobe-MolDyn:1999}.
All simulations are performed in the NV ensemble with periodic boundary conditions;
particle numbers are 16,000 in the MC simulations (4,000 for the larger densities
$\eta>0.6$ in 3D), and up to 256,000 in the MD 3D simulations.
The starting configuration is an ordered triangular or fcc lattice.

The pair correlation functions of the resulting configurations are found to be in agreement
with the Percus-Yevick approximation in the fluid phase \cite{Kalikmanov:2001}.
The mean square displacement of the particles between two sampled configurations
in the fluid phase is as large as the expectation for completely independent configurations,
indicating a sufficiently equilibrated system.
Away from the phase transition densities, both in 2D and 3D, the pair correlation functions
from MC and MD are in agreement.
\subsection{Averages of Anisotropy Indices}
In similar fashion as with the Poisson data, a periodic \Voronoi{} tessellation is computed
for each configuration of particles,
and Minkowski tensors and anisotropy indices are assigned to each particle.
Fig.~\ref{fig:hard-mu-of-beta} shows the averages of $\beta_\nu^{a,b}$ for hard disks
and hard spheres ensembles as a function of density.  It is seen that increased order
corresponds to more isotropic Voronoi cells in both hard disks and hard spheres.

We find MD and MC results to be in close agreement with each other except for in
the vicinity of the hard spheres coexistence region.
There, naturally, metastable states persist due to finite system sizes and simulation
times, and the state of the system depends heavily on
its history.  The metastable fluid branch can be traced even beyond the fluid/solid coexistence region
by choosing random initial conditions \cite{TomoakiNogawaHeatTransport}
instead of the usual fcc initial conditions
(see fig.~\ref{fig:hard-mu-of-beta}, inset in righthand side). Given enough
simulation time, it would eventually form fcc nuclei and crystallize.
The metastable fluid branch is distinctly more anisotropic on a local scale than the solid phase,
again confirming the rule of thumb that more ordered systems are more locally isotropic.

In both MC and MD simulations, a discontinuity in the average anisotropy index 
$\mu(\beta_\nu^{a,b})$ is observed at the hard spheres phase transition, with the solid phase being more
isotropic locally.  At the same time, the standard deviation of the distribution of $\beta_\nu^{a,b}$
drops discontinuously across the transition; in the solid phase, it goes
to zero almost linearly, $\sigma(\beta_\nu^{a,b})\propto (\eta-\wrmi\eta{max})$,
where $\wrmi\eta{max}$ is the maximum global packing fraction.

For large packing fractions, the system approaches the fcc state with
isotropic cells only ($\mu=1$, $\sigma=0$).  For small packing fractions, the Poisson
limit is recovered, and $\mu$, $\sigma$ attain their respective values for
the Poisson point process (bullets on the plot frame in figs.~\ref{fig:hard-mu-of-beta},
\ref{fig:hard-sigma-of-beta}).
Remarkably, tensors $W_2^{2,0}$ and $W_0^{2,0}$ yield
very similar anisotropy indices. No explanation for this data collapse
has been identified.  The shape of the distribution of $\beta$ values,
seen in fig.~\ref{fig:beta-distributions}, remains qualitatively similar
and unimodal
from the Poisson limit $\eta=0$ to the close-packed limit;
its average value has a discontinuity at the phase transition.
The development of the distribution function with varying packing fraction can be seen
in fig.~\ref{app:histo-hard-spheres} in the appendix.

The situation in the hard disks system is less clear, and the nature of the
phase transition has been the subject of continuous debate; current opinion favors
a two-step transition according to the Kosterlitz-Thouless-Halperin-Nelson-Young
scenario as opposed to a very weak
transition of first order (see \cite{BinderHardDisk:2002} and refs.~therein).

Mean anisotropy indices show a change of slope at the transition densities
(fig.~\ref{fig:hard-mu-of-beta}, left-hand).
In the transition region, we use systems of 64.000 disks in a hexagonal simulation cell with
periodic boundary conditions\footnote{A square box was also used. As expected,
it leads to artefacts for large packing fractions (starting from about
$\eta\approx 0.75$), and $\mu(\beta)=1, \sigma(\beta)=0$ is never attained as cells
are strained by the incommensurate simulation box.}.  Wo find no significant differences
between systems of 16.000 and 64.000 disks, given enough relaxation time.
The $\mu(\beta)$ curves in fig.~\ref{fig:hard-mu-of-beta}, left-hand side, and 
second moments $\mu(\beta^2)$ behave linearly as prescribed by the lever rule,
with coexistence densities of $\approx$0.70 and 0.72. Binder et al.~\cite{BinderHardDisk:2002}
report closer values of 0.706 and 0.718. Full equilibration
of the hard disks system in the transition region, however, remains difficult even with
more advanced algorithms than the ones used in the present study \cite{PhysRevE.80.056704}.

\subsection{Correlations among Anisotropy Indices}
We find that the four (in 2D; six in 3D) different anisotropy indices display
a qualitatively similar behavior.  This shows that the anisotropy of the \Voronoi{} cells
is a generic feature and does not depend on the particular aspect of anisotropy being
analyzed.  Shapes can easily be constructed where the
different Minkowski tensors yield widely different anisotropy indices.  However,
such shapes do not generally occur in a \Voronoi{} tessellation.

This motivates the use of a single anisotropy index instead of the full set
of Minkowski tensors.  To justify this, we calculate the
correlation coefficients $\corr(X,Y):=\cov(X,Y)/(\sigma(X)\sigma(Y))$,
$\cov(X,Y):=\mu((X-\mu(X))(Y-\mu(Y)))$, of the anisotropy indices;
most pairings of $\beta_\nu^{a,b}$ are correlated with $\corr > 0.8$
over the whole packing fraction range of the
hard spheres ensemble.  Only $\corr(\beta_0^{2,0},\beta_2^{0,2})$ gets
as low as $0.7$ for the smallest packing fractions. The same
is true for the hard disks ensemble, $\corr>0.8$ for most pairings; only
$\corr(\beta_0^{2,0},\beta_1^{0,2})$ gets as low as 0.65 for the most dilute systems.
It is thus justified,
for the anisotropy analysis of the hard spheres ensemble, to focus on a single
convenient anisotropy index, for example the translation invariant $\beta_1^{0,2}$.
In the remainder of the article, we drop the indices $\nu, a, b$
off the $\beta$.

\subsection{Comparison to Isoperimetric Ratio}
Measures of asphericity have been previously used,
e.\,g.~the isoperimetric ratio (also called \emph{shape factor}
\cite{PhysRevLett.94.040601,PhysRevLett.96.258001}), defined by
$\zeta := (W_1)^2 (W_0)^{-1} / \omega_2$ in two dimensions,
and $\zeta:=(W_1)^3 (W_0)^{-2} / \omega_3$ in three dimensions,
with the volume $\omega_D$ of the $D$-dimensional unit sphere.
Starr et al.~use an asphericity index based on
closest neighbors distance in the \Voronoi{} tessellation
\cite{StarrSastryDouglasGlotzer:2002}.

The distinction between asphericity, meaning deviations of the shape from
a sphere, and anisotropy, meaning that the object has different spatial extension
or surface area in different spatial directions, is subtle. Clearly, a sphere's
isoperimetric ratio can be increased by applying an undulation to the
bounding surface or by deforming it into a cube.  Both leave the anisotropy
unchanged.  On the other hand, $\zeta$ also changes when the sphere is deformed
into an ellipsoid, becoming anisotropic.  The anisotropy indices $\beta$
clearly distinguish between asphericity and anisotropy.
Also, the Minkowski tensors not only provide
a convenient anisotropy index, but also identify explicitly, by means of the eigenvectors,
the preferred directions for each cell.
Accordingly, we find that $\zeta$ is less correlated with the $\beta$ indices than the set
of $\beta$ among themselves; $\corr(\zeta, \beta_\nu^{a,b})$ ranges between $-0.6$ and $-0.2$ for
all $\beta_\nu^{a,b}$ both in two and three dimensions, meaning that the Minkowski tensors
probe a different aspect of the cell shapes than the isoperimetric ratio alone.

It is noteworthy that the isoperimetric ratio follows a double-peak distribution
in the hard disks system \cite{PhysRevLett.94.040601}.
The distribution of the $\beta$ indices, however, is unimodal, see fig.~\ref{fig:beta-distributions};
it changes continuously across the solid/fluid phase transition.

\section{Lennard-Jones fluid}
\label{sec:lennard-jones}%
\begin{figure}
\centering
\includegraphics[scale=1.3]{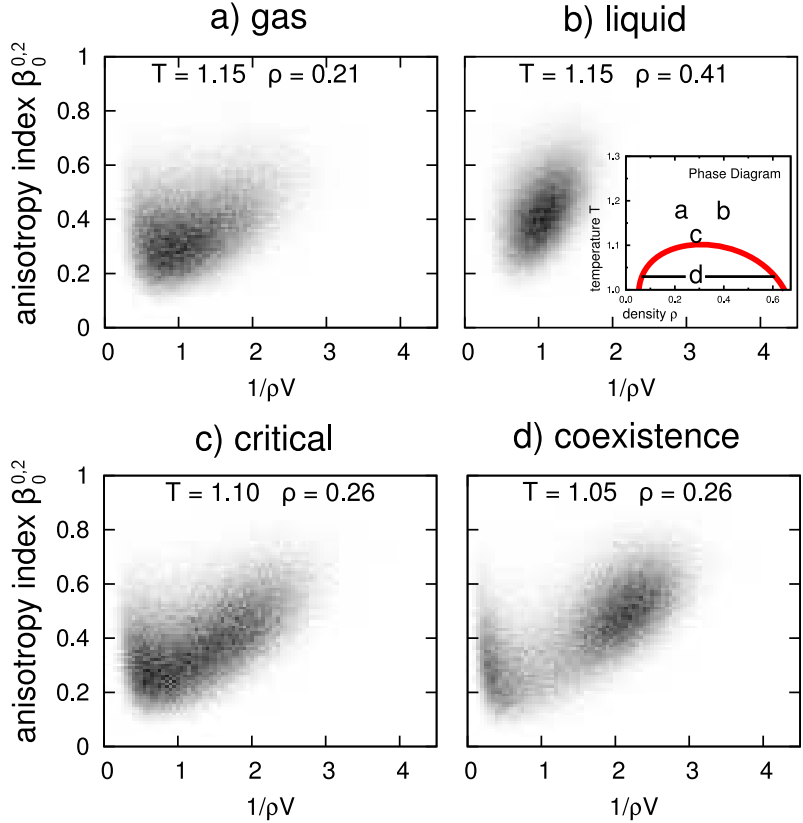}

\caption{Histograms of the \Voronoi{} cell anisotropy and density at four points in
the Lennard-Jones phase diagram (sketched in inset):  a) vapor state
b) liquid state c) critical composition d) coexistence of liquid and vapor.
The systems consist of 100.000 particles each.  The $x$ axis shows a local normalized density
given by the reciprocal cell volume $V$ in units of the global density $\rho$, while
the $y$ axis shows the anisotropy index $\beta_0^{2,0}$.  A more complete survey of the phase
diagram is found in the appendix (fig.~\ref{app:histos-lj-phase-diagram}).}
\label{fig:lj-beta-histos}
\end{figure}

Configurations of Lennard-Jones (LJ) fluids with 100.000 particles in periodic boundary
conditions are generated using a parallel molecular dynamics code
\cite{DissertationChrisGoll:2010,FrenkelUnderstandingMolSim:2002}.
The LJ fluid is a true thermal system, the phase diagram is two-dimensional,
and includes a critical point.  The precise location of the critical point depends on
the numerical treatment of the LJ potential, in particular the cutoff radius and
cutoff corrections \cite{trokhymchuk:8510}.  Here, a cutoff of $2.5 r_0$ is chosen.

For the pointlike LJ particles, we consider, as previously in fig.~\ref{fig:first-page-picture},
the local normalized (particle number) density $d^{(i)}=1/(\rho V^{(i)})$,
where $\rho$ is the global particle number density, and the volume  $V^{(i)}:=W_0(\bodyI{})$ 
is the volume of \Voronoi{}
cell $\bodyI{}$.  Thus, $\mu\bigl( 1/d^{(i)}\bigr) = 1$.
Fig.~\ref{fig:lj-beta-histos} shows two-parameter histograms of the densities $d^{(i)}$
and anisotropy indices $\beta_0^{2,0}(\bodyI{})$ of Voronoi cells in LJ configurations.
Configurations from four points in the LJ phase diagram are shown, and the corresponding
locations indicated in the phase diagram sketch (inset).
In the high-temperature limit, above the critical point, we find a continuous
change of the probablity distribution from a triangular shape (subfigure a) 
as is also found for the ideal gas (fig.~\ref{fig:first-page-picture}) to a ellipsoidal shape (subfigure b)
as is found in the hard spheres system (also seen in fig.~\ref{fig:first-page-picture}).
Below the critical point, coexisting liquid and vapor phases can be clearly distinguished
as two components of the probability distribution (subfigure d).
These components grow closer as the temperature is increased, and eventually merge
at the critical point (subfigure c).

\section{Robustness of the Minkowski Tensors}
\label{sec:misc}%
\begin{figure}
\centering
\includegraphics{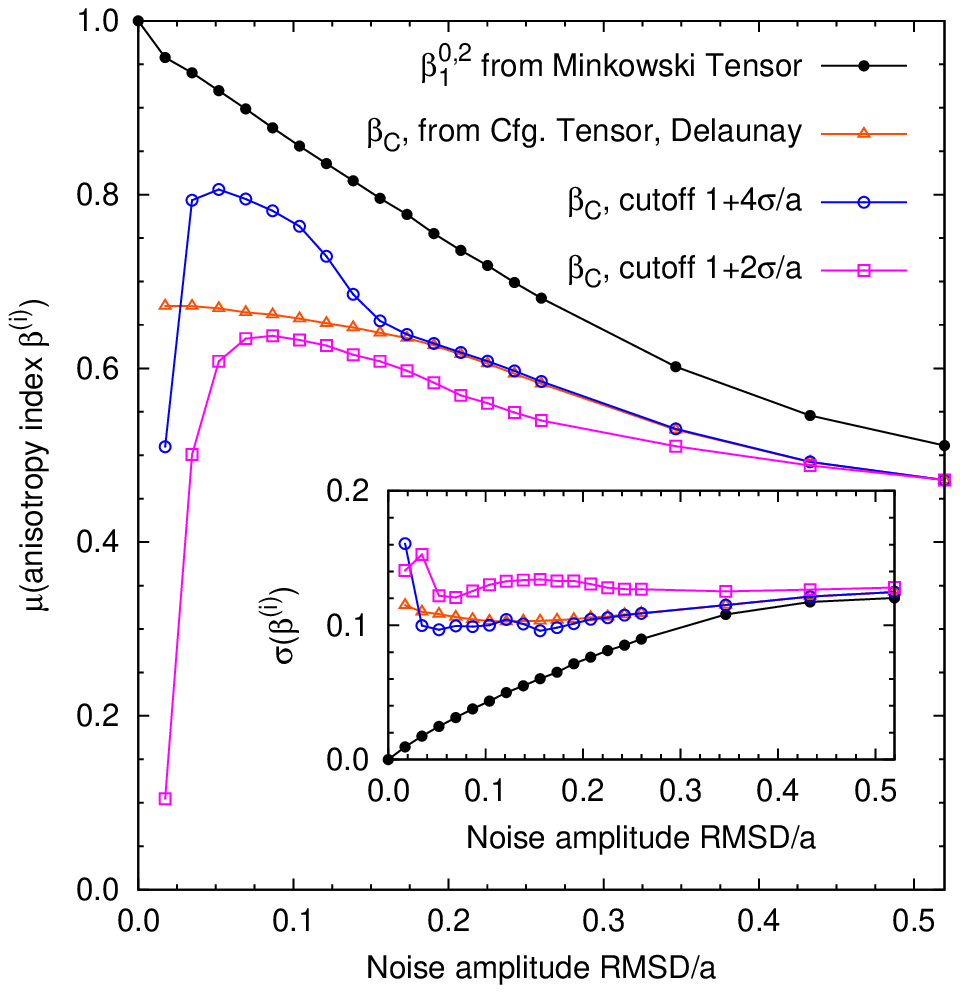}
\includegraphics{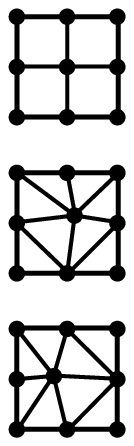}

\caption{\emph{Left:} Anisotropy index for $W_1^{0,2}(\bodyI{})$ and Edwards' configurational tensor $C^{(i)}$,
applied to a fcc Einstein solid.  The noise amplitude is quantified by the
root mean square displacement $\operatorname{RMSD}$ of the germs  from their ideal lattice sites.
The Minkowski tensor $W_1^{0,2}(\bodyI{})$ is computed from the \Voronoi{} tessellation of
the lattice, $C^{(i)}$ is computed with the Delaunay neighborhood and distance cutoff
neighborhoods of the order of the noise amplitude.  It is seen that,
even for vanishingly small levels of noise, $C^{(i)}$ of a single lattice site
is not an isotropic tensor, while $W_1^{0,2}(\bodyI{})$ is.
Tensor isotropy is defined via the ratio of eigenvalues, as in eq.~\eqref{def:beta}.
The $x$ axis and cutoffs are
in units of the fcc nearest neighbor distance $a$.

\emph{Right:}  Illustration of point configuration with a degenerate Delaunay triangulation.
A small perturbation (not drawn to scale)
lifts the degeneracy and breaks the isotropic 4-simplices
into anisotropic triangles.}
\label{fig:edwards-is-not-robust}
\end{figure}
A simple model of a thermal solid is defined by putting the germs on an ideal lattice,
and allowing each germ to deviate by a small distance $\vec \varepsilon^{\;(i)}$ from its ideal site,
as in the Einstein solid.
We generated a point configuration by drawing, independently for each lattice site,
the random displacement vector $\vec \varepsilon^{\;(i)}$ from a three-dimensional
Gaussian distribution.  The set of germs is then $\{ \vec g^{\;(i)} = \vec l^{\;(i)} + \vec \varepsilon^{\;(i)}\}$,
where $l^{(i)}$ are the fcc lattice sites.
The noise amplitude is given by the root mean square displacement (RMSD) of the
particles, $\operatorname{RMSD} = \sqrt{\mu(|\vec \varepsilon^{\;(i)}|^2)}$; since
the expectation value of $\vec \varepsilon^{\;(i)}$ vanishes, $\operatorname{RMSD} = \sigma(\vec \varepsilon^{\;(i)})$.
Increasing the noise amplitude eventually destroys lattice order
und yields a Poisson point pattern.  This model neglects correlations between
displacements of different particles that necessarily exist in a real solid, and does allow
for overlap if the particles have finite radius.

In fig.~\ref{fig:edwards-is-not-robust}, as the noise amplitude is increased, 
$\mu(\beta_1^{0,2})$ decreases continuously from unity,
similarly as it does when decreasing packing fraction $\eta$ from the maximum
density state in the hard spheres model (fig.~\ref{fig:hard-mu-of-beta}).
In the limit of strong noise, the anisotropy index of the Poisson point pattern
is recovered ($\mu(\beta_1^{0,2})\rightarrow 0.457$).

\subsection{Robustness against Noise}
The Minkowski tensors of Voronoi cells vary continuously when germs as dislocated.
The topology of the Voronoi
tessellation may change; however, the shape of the Voronoi cells $K^{(i)}$ is a
continuous function of the germ point locations.  For example, it is possible
that a small dislocation of a germ creates additional facets in the \Voronoi{}
tessellation; however, since these new faces are very small, and their contribution
to the tensor of the cell is weighted with their area, only a small change in
the tensor is effected.

This continuity property is not necessarily guaranteed for other,
similarly defined shape measures.  For example,
Edwards et al.~\cite{Edwards2001162,EdwardsBrujicMakse:2005}
defined a configurational tensor
\begin{equation}
    {C^{(i)}} := \sum_{j\in\operatorname{NN}(i)}
        (\vec r^{\,(i)}-\vec r^{\,(j)}) \otimes (\vec r^{\,(i)}-\vec r^{\,(j)})
\end{equation}
with some suitable definition for the set NN of nearest neighbors.
NN may be defined using the Delaunay triangulation, or a fixed cutoff radius,
or physical contacts.
If, under a small dislocation of a germ, the neighborhood NN changes,
a comparatively large change of the tensor results, because the bond vector
$\vec r^{\;(i)}-\vec r^{\;(j)}$ to the lost or newly gained neighbor tends to be larger
than the other bond vectors. Consequently, $C^{(i)}$ changes discontinuously\footnote{%
Note that Edwards et al.~use $C^{(i)}$ to describe 
physical contact points in granular matter, and for that purpose, $C^{(i)}$ must change
discontinuously.  However, as a consequence, $C^{(i)}$ is not suitable to characterize local
anisotropy in a robust way.}.

The discontinuous behavior of $C^{(i)}$ is especially pronounced for
near-degenerate configurations, for example the fcc lattice with
vanishingly small noise amplitude.  As fig.~\ref{fig:edwards-is-not-robust} shows,
$C^{(i)}$ is not an isotropic tensor at zero noise, even though a fcc lattice
site has 12 symmetrically arranged nearest neighbors.  This is due to numerical
roundoff errors.  For simplicity, we discuss the two-dimensional case, which exhibits
the same conceptual problem.  Consider the perfect square lattice in
fig.~\ref{fig:edwards-is-not-robust}.  The Delaunay simplices are not, as in a general
point pattern, triangles, but degenerate to squares. $C^{(i)}$ computed from
such simplices is an isotropic tensor. However, tiny amounts of numerical inaccuracy
or noise lift the degeneracy for some of the squares and cause the squares
to break up into triangles
\cite{BarberDobkinHuhdanpaa:1996}.  $C^{(i)}$ computed from the triangles is no longer
an isotropic tensor.  Due to the same effect in three dimensions, $C^{(i)}$ with
the Delaunay neighborhood is not a good measure of anisotropy.

Note that the Minkowski tensor $W_1^{0,2}$ is similar to the Edwards tensor as
the normal directions of \Voronoi{} facets are precisely the directions of the
bond vectors via the \Voronoi{}--Delaunay duality. However, the weighting with the
facet area in eq.~\eqref{eq:w102sumrepr} ensures that the tensor is a
continuous function of the germ point locations.

\subsection{Discrimination of Einstein Solid and Hard Spheres Equilibrium Solid}
As shown above, the mean anisotropy index $\mu(\beta)$ of the Einstein solid
decreases with increasing noise amplitude, similar to the hard spheres solid.
However, the two point patterns are clearly structurally different, as the hard spheres
system includes correlations between individual particles.

\begin{figure}
\centering
\includegraphics{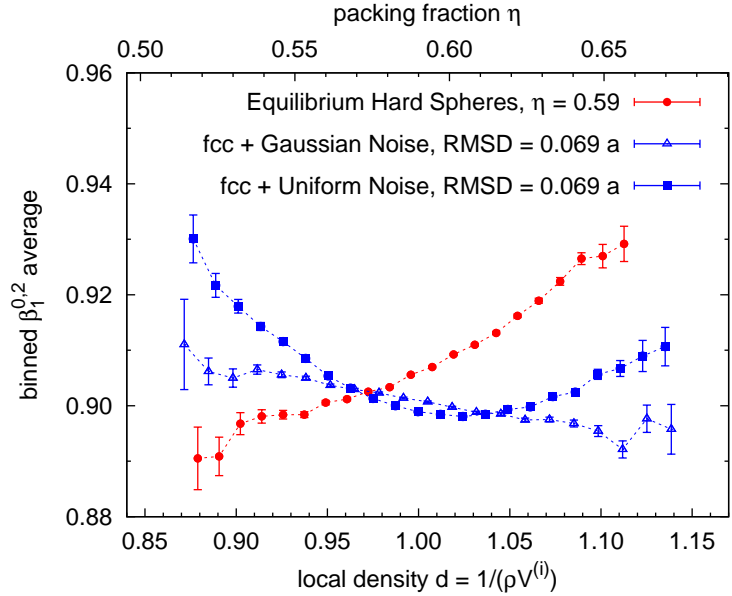}

\caption{Comparison of average anisotropy index $\beta_1^{0,2}$ as a function of
local normalized densities $d$.  The \Voronoi{} cells are binned in density bins,
and an average of their anisotropy indices is computed separately for each bin.
The error bars are estimated from standard deviations in each bin.
The plot shows that the correlation of anisotropy index and cell volume distinguishes
between the Einstein solid and the equilibrium hard spheres structure
(in the solid phase), even though parameters were chose such that 
the global averages $\mu(\beta_1^{0,2})$ are equal.}
\label{fig:difference-true-solid-model-solid}
\end{figure}

To see this, it is again useful to study the correlation between
local normalized density $d$ and local anisotropy $\beta$.
Figure \ref{fig:difference-true-solid-model-solid} shows that $d^{(i)}$ is negatively
correlated with $\beta(\bodyI{})$ for the Einstein solid (blue dashed line), while the two variables are
positively correlated for the hard spheres solid (red solid line).
Positive correlation is the rule for interacting systems such as hard spheres,
and has also been found for jammed bead packs \cite{NonScience:2010}, while
a negative correlation has been found for foam models.

This can be understood
considering that, for the hard spheres solid, a higher density implies that the cell shape
is closer to the shape of the particle since particles cannot overlap. On the other hand, 
for the Einstein solid, no minimum distance is enforced, and high densities are 
the result of one germ intruding into the domain of another, creating two 
oblate \Voronoi{} cells.

\section*{Conclusions}
This study shows how Minkowski tensors can be used to characterize
point patterns, and in particular their local anisotropy.
A simple, linear algorithm to compute the shape tensors of a polyhedron was given elsewhere
\cite{SchroederTurkKapferBreidenbachBeisbartMecke:2009,MickelBreidenbachMeckeSchroederTurk:2008}.

Particle configurations from 
a number of fluid models are analyzed using the Minkowski tensor framework,
and reference values for these idealized systems are given.
In each of the systems, the Minkowski tensor analysis is able
to quantify the morphological properties of the system in a robust
way.  The continuity properties guarantee that small shifts in the germ
positions do not influence the result in an undue manner, as is the
case for a na\"ively defined ``bond tensor''. In particular, $W_1^{0,2}$
is found to include the correct weighting prefactors of the bond vectors
to ensure continuity.

The anisotropy analysis is found to be sensitive to the phase behavior of the fluids,
and the average anisotropy index may be used to distinguish collective states of the fluid.
More research is needed in order to link order parameters based
on Minkowski tensors with more established order parameters like the bond-angle 
\cite{PhysRevE.55.6855} order parameter.  For the hard spheres system,
exact relations between the equations of state and the geometry of the available space,
in particular its surface area and volume,
are known to exist \cite{Speedy:1980,SpeedyReiss:1991,Corti:1999,Sastry:1998}.
Similar relations are not yet known for Minkowski tensors of available space.
However, a density functional theory based on Minkowski tensors of convex particles 
has been developed, and provides accurate approximations for the equations of state
\cite{Rosenfeld:1989,HansenGoos:2009,HansenGoos:2010}.
Finally, the application of the Minkowski tensors to a comprehensive set of experimental
data is desirable.  For the latter purpose, the Minkowski tensor software may be
freely downloaded from the institute homepage at
\begin{center}
\texttt{http://www.theorie1.physik.uni-erlangen.de/}
\end{center}

\section*{Acknowledgments}
We thank Tom Truskett and Roland Roth for helpful comments.  Dominik Krengel contributed
some analysis scripts. GEST and SCK acknowledge financial support 
by the Deutsche Forschungsgemeinschaft under grant SCHR 1148/2-1.

\section*{References}
\bibliography{./LITERATURE/literatur}
\bibliographystyle{unsrt.bst}
\begin{figure}[b]
\centering
\includegraphics{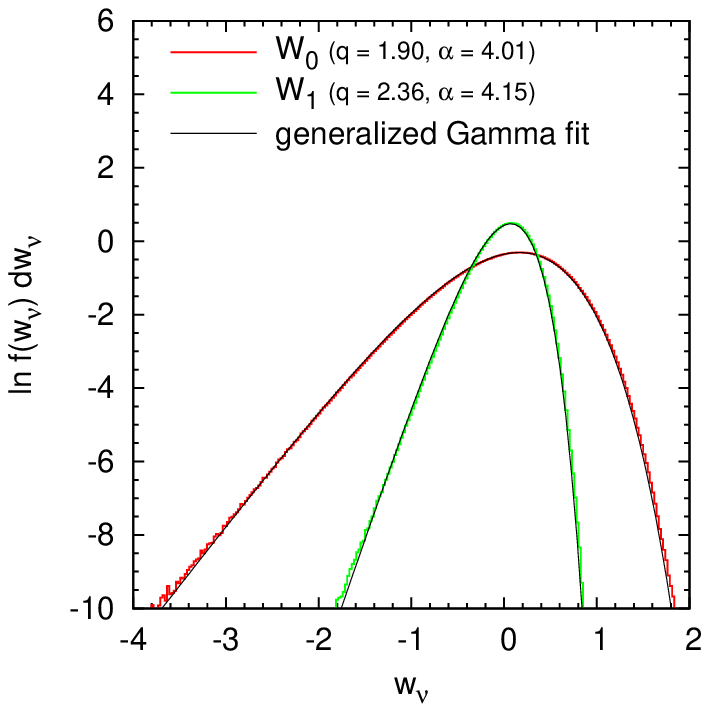}
\includegraphics{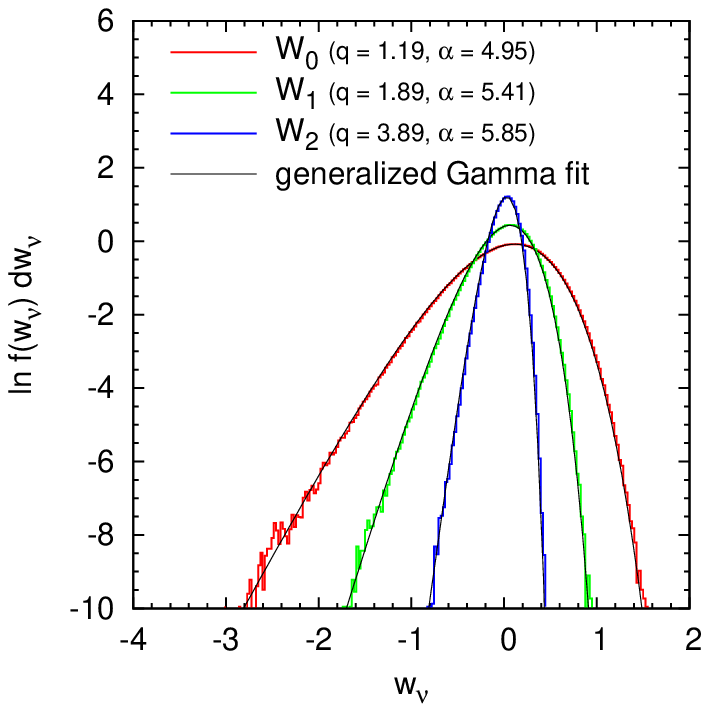}

\caption{Distribution of scalar cell properties in a Poisson-Voronoi tessellation:
The probability density is logarithmized, and its argument replaced by the
reduced quantity $w_\nu:=\ln W_\nu - \mu(\ln W_\nu)$.  In this plot, a log-normal
distribution would appear as a parabola with the apex on the $w=0$ line.
The data is generated from $5\times10^7$ (2D) and $10^7$ (3D) random Voronoi cells.
Generalized Gamma distributions $\propto (W_\nu)^{q(\alpha-1)} \exp(-(W_\nu/\theta)^q)$ give good fits
with the parameters specified in the plot key.  ($\theta$ sets the length scale and is fixed by the 
normalization condition $\mu(w_\nu)=0$.)

%FIT IN THE <W>=1 NORMALIZATION
%4.08493316910303 0.327871552056869 1.07618752739006
%4.19888464980202 0.342565858926597 2.33746174069682
%5.0570589978564 0.251107613918575 1.17402039793225
%5.54839655349235 0.231287990776268 1.85708132847746
%5.92196465633955 0.234923655751431 3.86078504826622
%
%FIT IN THE <w>=0 NORMALIZATION
%4.00865083387409 0.396183527605484 1.09221625379751
%4.14543085865665 0.376492519762817 2.36094114555981
%4.9471200855683 0.289739281754529 1.19173492196527
%5.41440771980872 0.255212371077006 1.88696056911649
%5.85294783274318 0.229770084027769 3.89352124116208
%
%\begin{center}
%\begin{tabular}{llll}
%     & Fct.   & $q$   & $\alpha$ \\[.3em]
%$D=2$&$W_0$   & 1.09  & 4.01 \\
%  &   $W_1$   & 2.36  & 4.15 \\[.3em]
%$D=3$&$W_0$   & 1.19  & 4.95 \\
%  &   $W_1$   & 1.89  & 5.41 \\
%  &   $W_2$   & 3.89  & 5.85 
%\end{tabular}
%\end{center}
}
\label{fig:scalar-prop-distros}
\end{figure}

\newcommand\histbox[2]{\parbox{.18\linewidth}{\centering\includegraphics[width=\linewidth]{#1} $\eta = #2$}}
\begin{figure}
\centering
\histbox{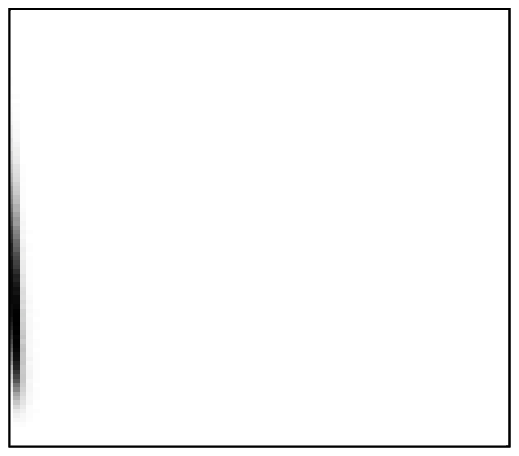}{0.01}
\histbox{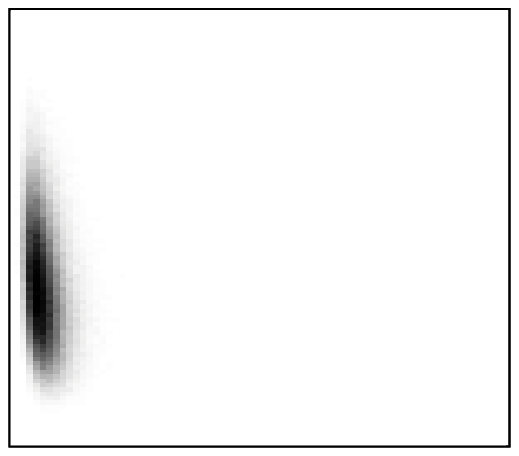}{0.05}
\histbox{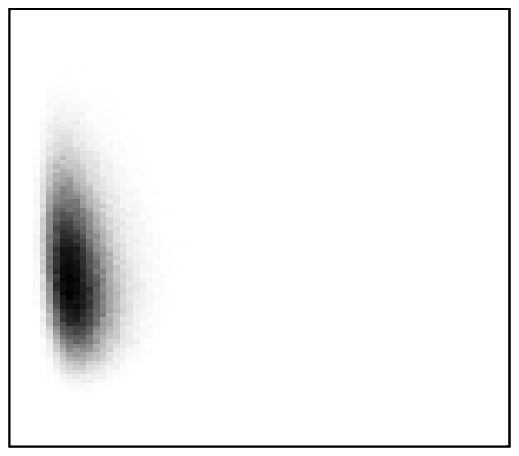}{0.10}
\histbox{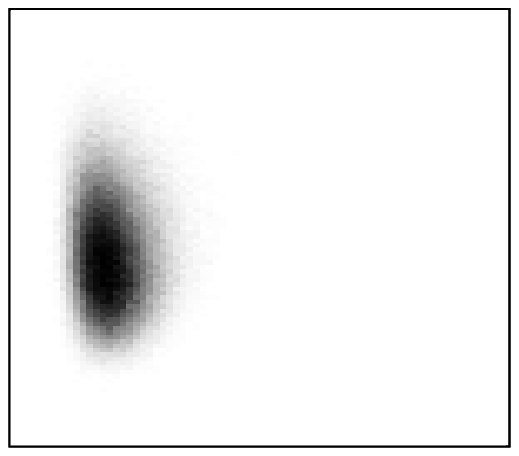}{0.15}
\histbox{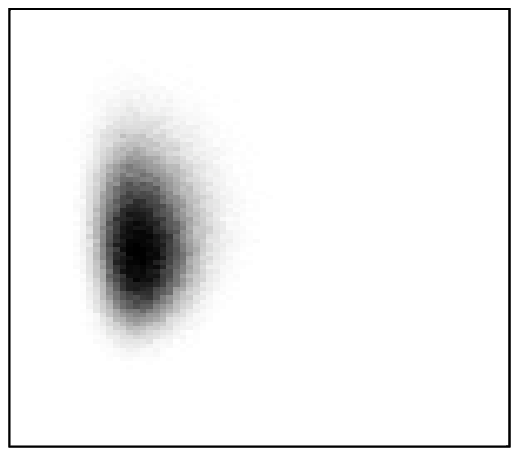}{0.20}
\vspace{.5em}

\noindent%
\histbox{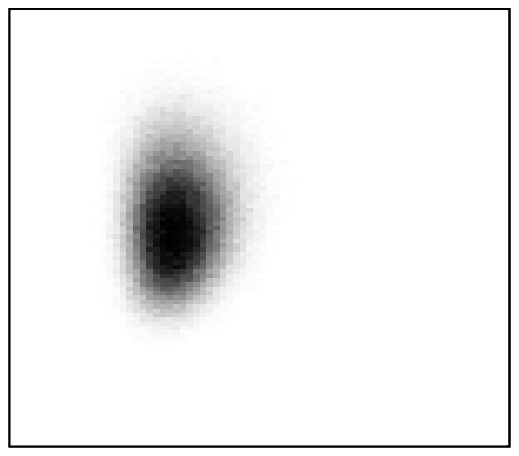}{0.25}
\histbox{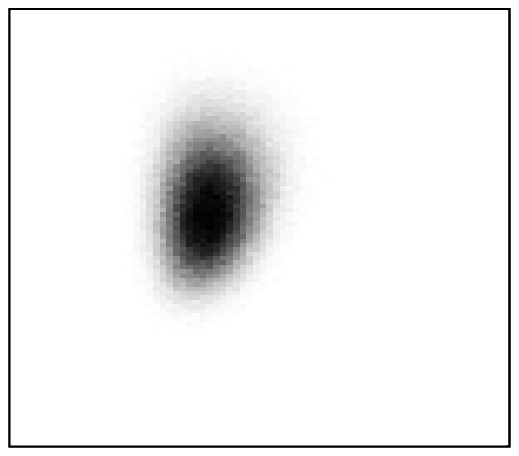}{0.30}
\histbox{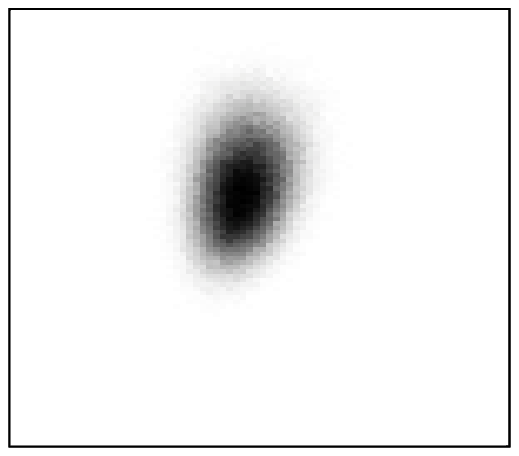}{0.35}
\histbox{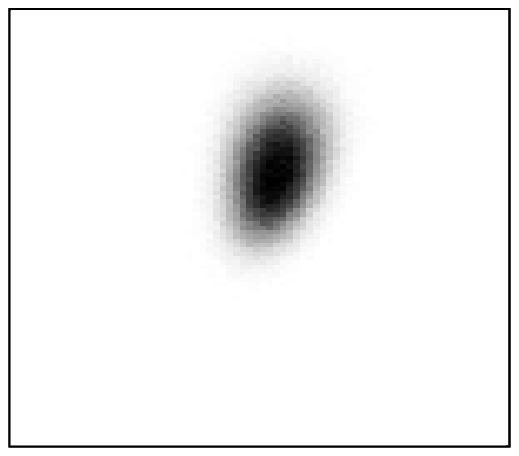}{0.40}
\histbox{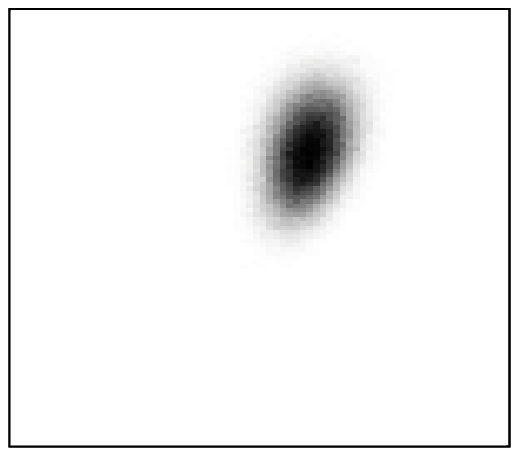}{0.45}
\vspace{.5em}

\noindent%
\histbox{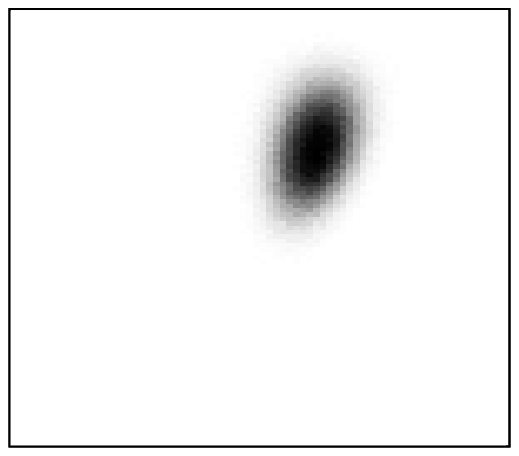}{0.46}
\histbox{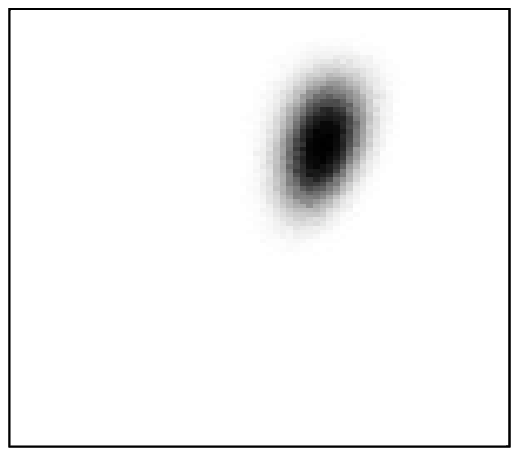}{0.47}
\histbox{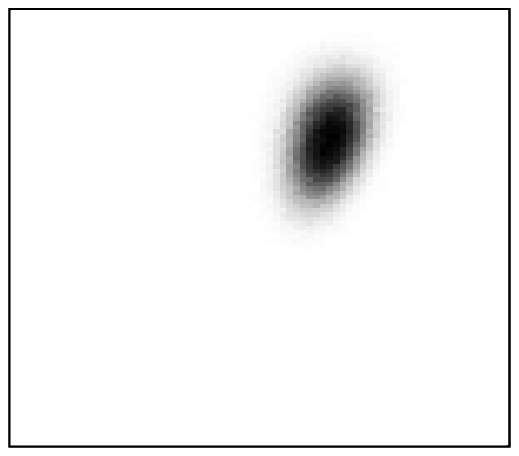}{0.48}
\histbox{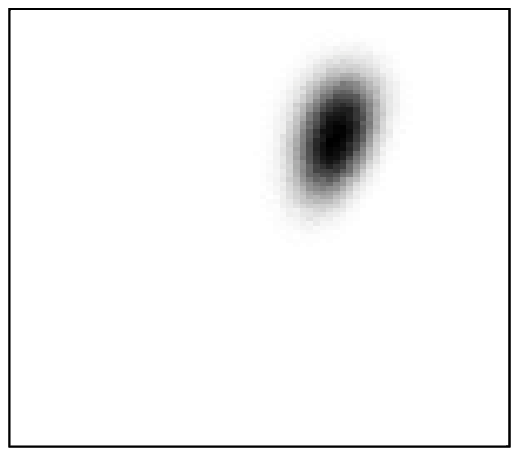}{0.49}
\histbox{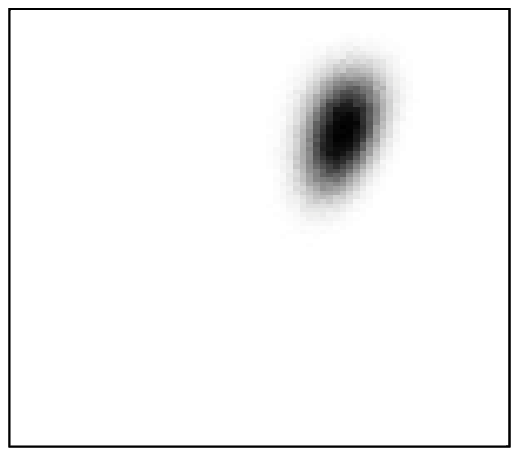}{0.50}
\vspace{.5em}

\noindent%
\histbox{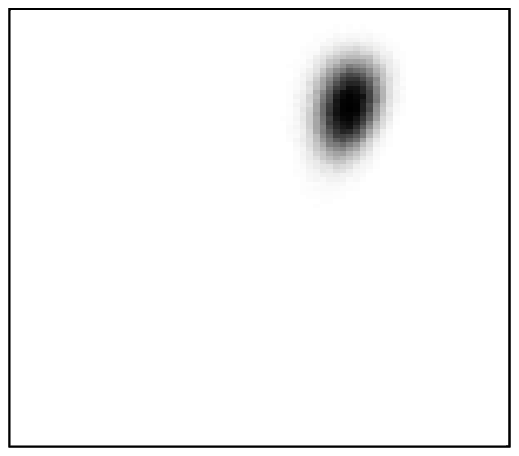}{0.51}
\histbox{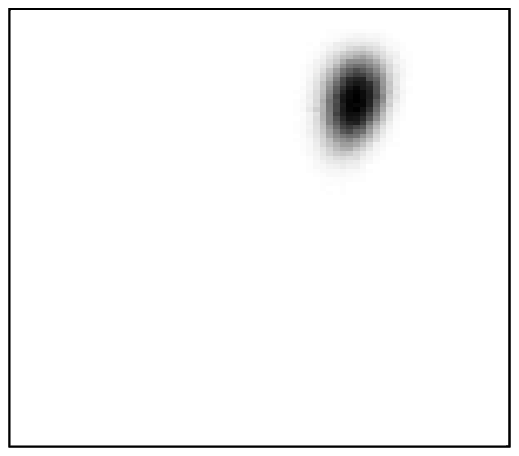}{0.52}
\histbox{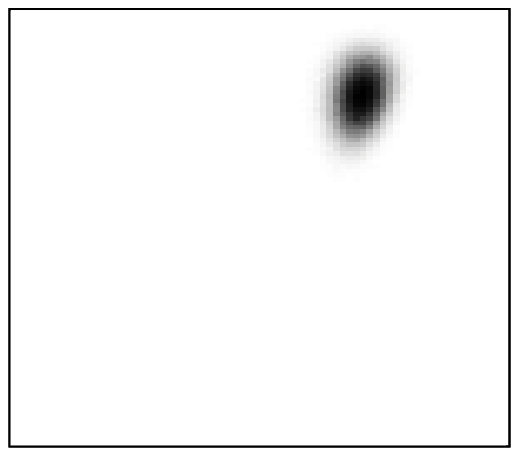}{0.53}
\histbox{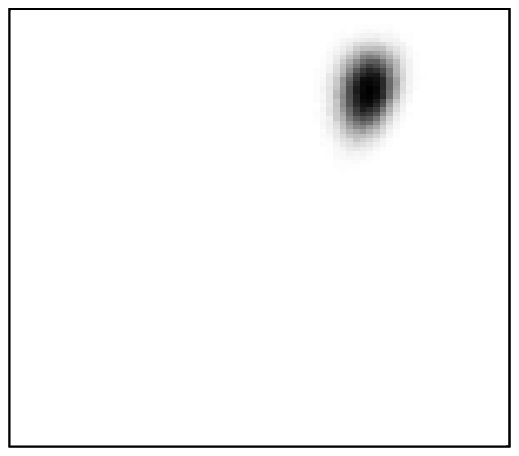}{0.54}
\histbox{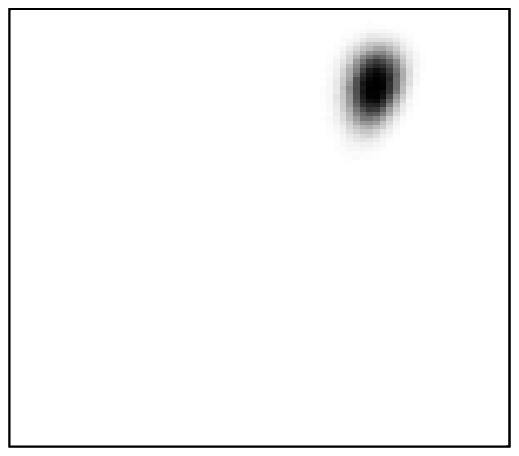}{0.55}
\vspace{.5em}

\noindent%
\histbox{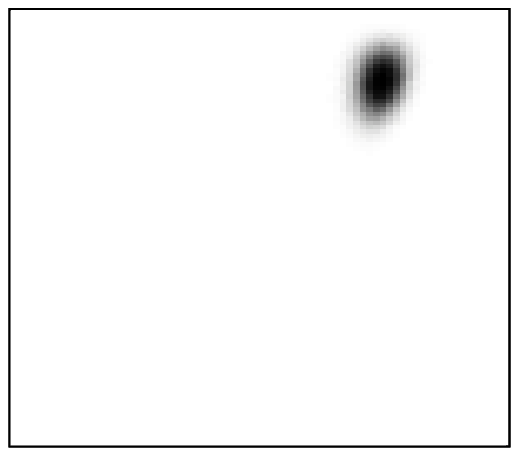}{0.56}
\histbox{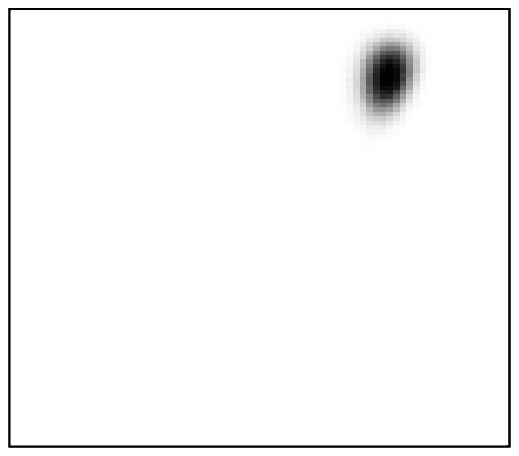}{0.57}
\histbox{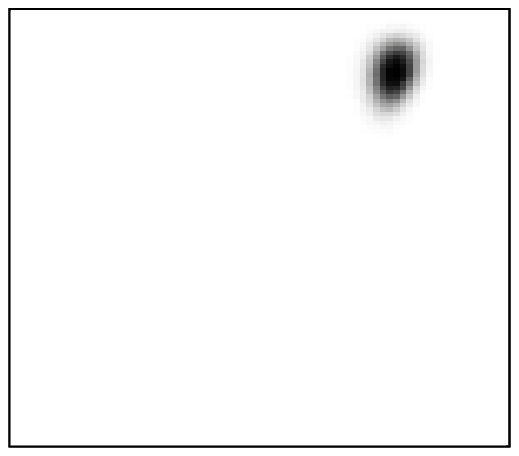}{0.58}
\histbox{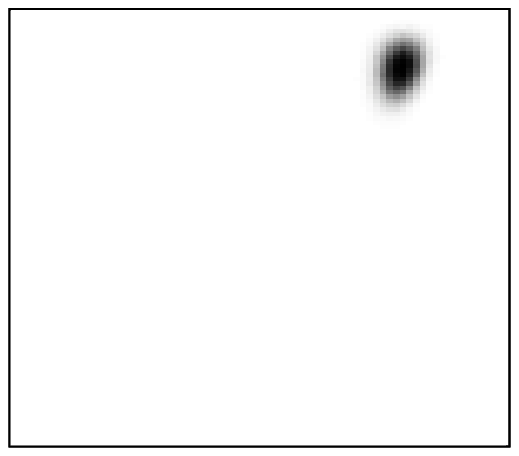}{0.59}
\histbox{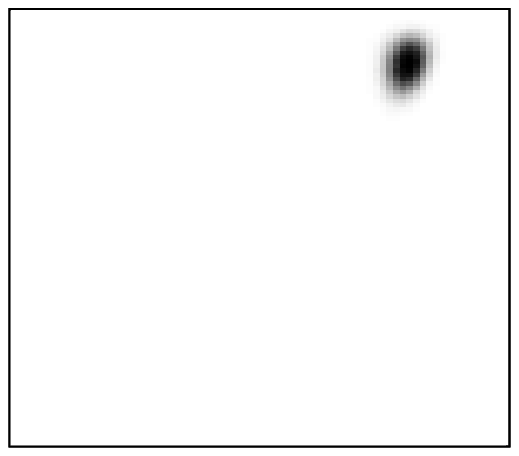}{0.60}
\vspace{.5em}

\noindent%
\histbox{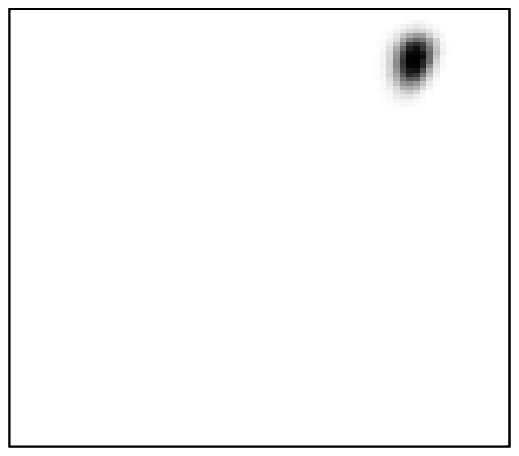}{0.61}
\histbox{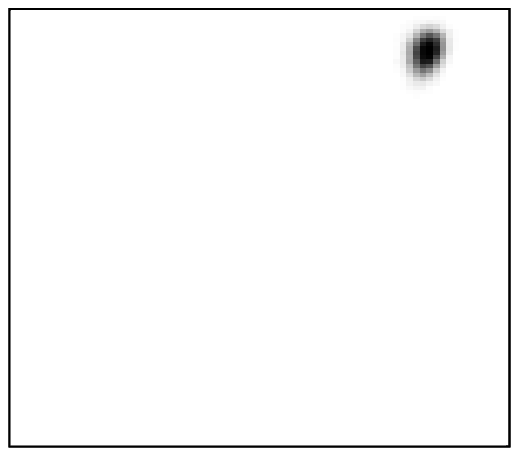}{0.63}
\histbox{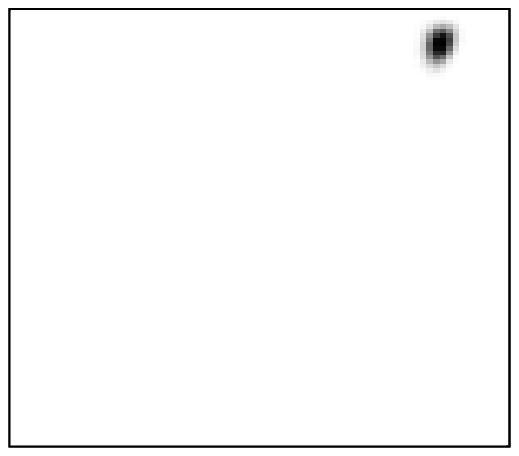}{0.65}
\histbox{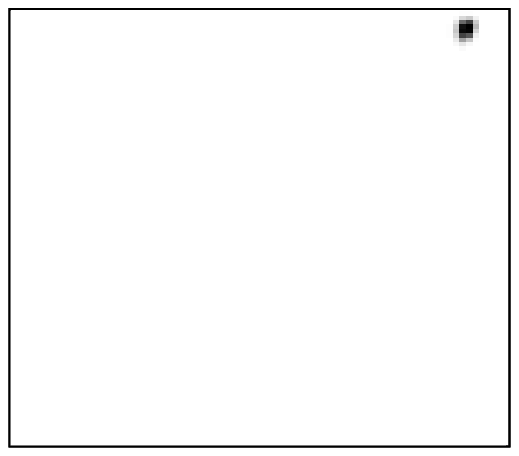}{0.67}
\histbox{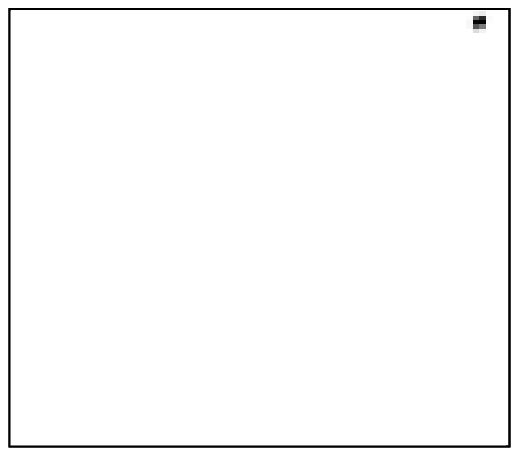}{0.71}

\caption{Histograms of anisotropy index $\beta_0^{2,0}$ and packing fraction $\eta$
of Voronoi cells in the hard spheres ensemble.
The $x$ axis of each box shows the packing fraction $\eta^{(i)}\in[0; 0.75]$ of individual cells,
the $y$ axis shows the anisotropy parameter $\beta_0^{2,0}(K^{(i)})\in[0; 1]$.
Darker gray corresponds to larger probablity.  The distribution
changes characteristically from a triangular shape ($\eta<0.2$) to
a ellipsoidal distribution for higher densities. In the limit $\eta\rightarrow\pi/\sqrt{18}\approx 0.74$,
the distribution becomes a Dirac distribution $f(\eta,\beta)=\delta(\eta-\pi/\sqrt{18})\delta(1-\beta)$.
}
\label{app:histo-hard-spheres}
\end{figure}
\renewcommand\histbox[2]{\parbox[b]{.18\linewidth}{\centering\includegraphics[width=\linewidth]{#1}}}
\begin{figure}
\centering
%\histbox{AppLennardJones/hist2D_fluid_w020_n_5000000_t_1_txt}{}
\parbox[b]{.6cm}{\rotatebox{90}{$T=1.50$}}
\histbox{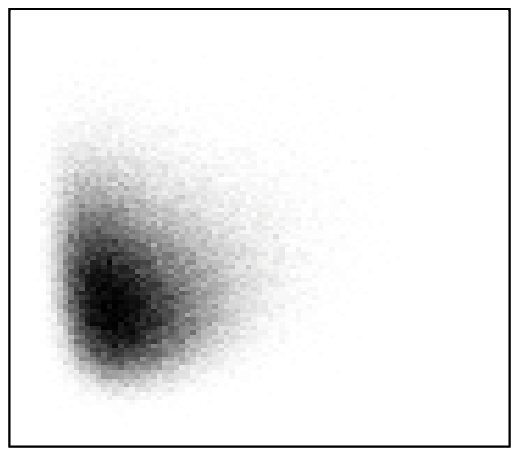}{}
\histbox{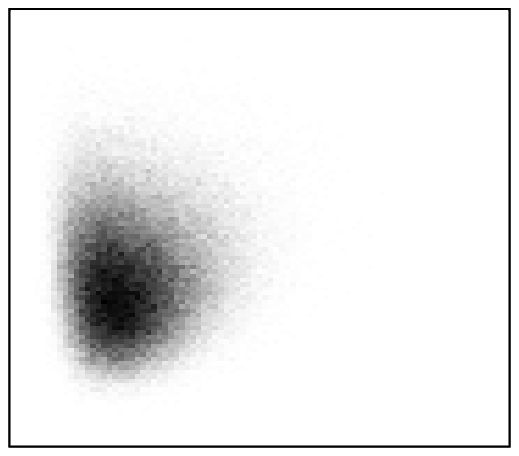}{}
\histbox{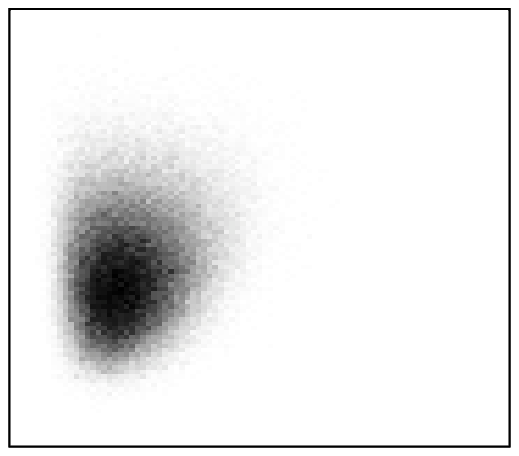}{}
\histbox{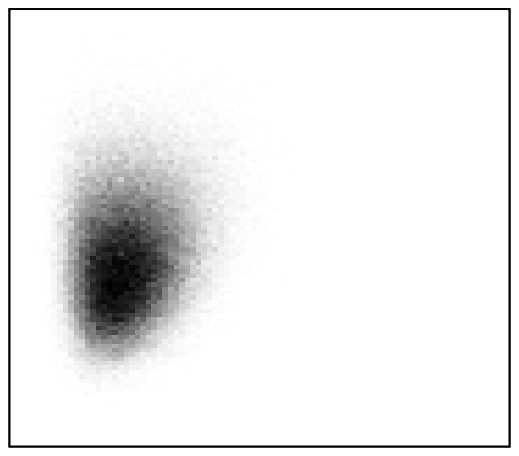}{}
\histbox{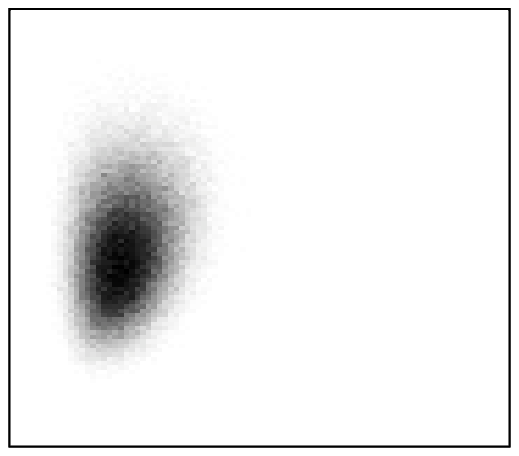}{}
\vspace{.5em}

%\noindent
%\histbox{AppLennardJones/hist2D_lj_w020_T_1_2_rho_0_26_txt}{}
%\histbox{AppLennardJones/hist2D_lj_w020_T_1_2_rho_0_31_txt}{}
%\histbox{AppLennardJones/hist2D_lj_w020_T_1_2_rho_0_36_txt}{}
%\histbox{AppLennardJones/hist2D_lj_w020_T_1_2_rho_0_41_txt}{}
%\vspace{.5em}

\noindent
\parbox[b]{.6cm}{\rotatebox{90}{$T=1.15$}}
\histbox{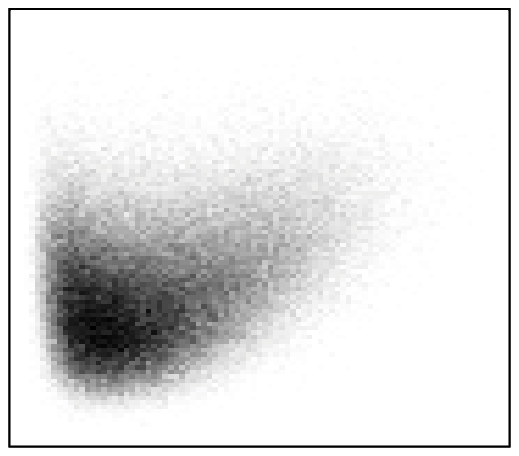}{}
\histbox{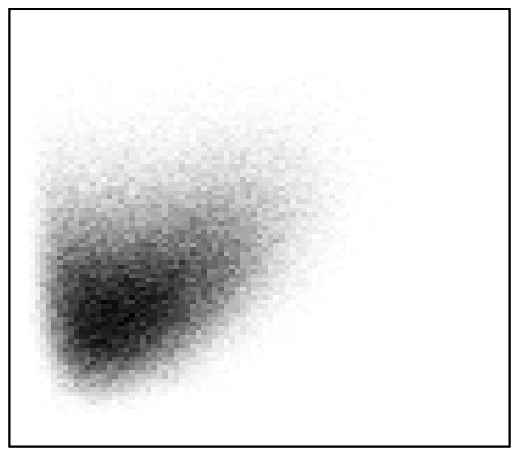}{}
\histbox{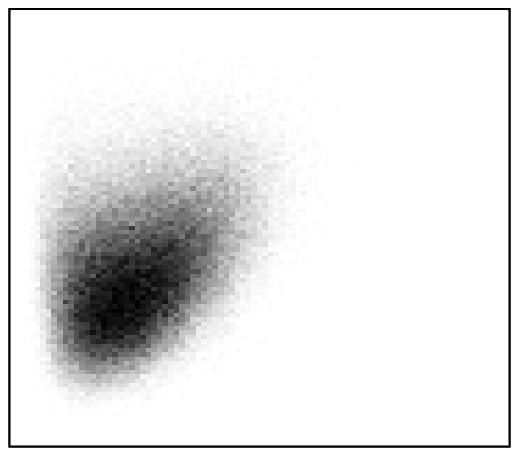}{}
\histbox{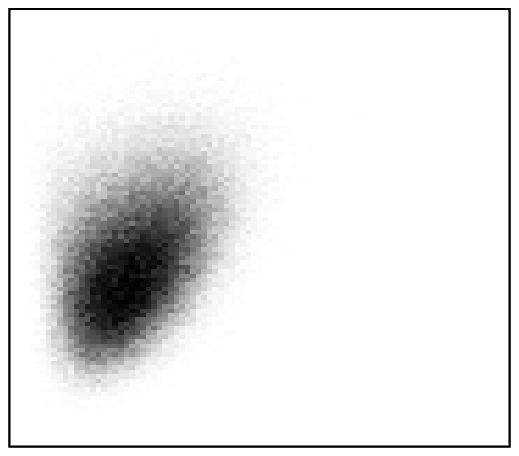}{}
\histbox{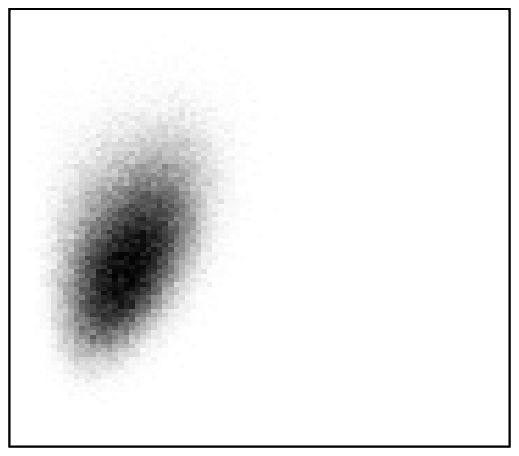}{}
\vspace{.5em}

\noindent
\parbox[b]{.6cm}{\rotatebox{90}{$T=1.10$}}
\histbox{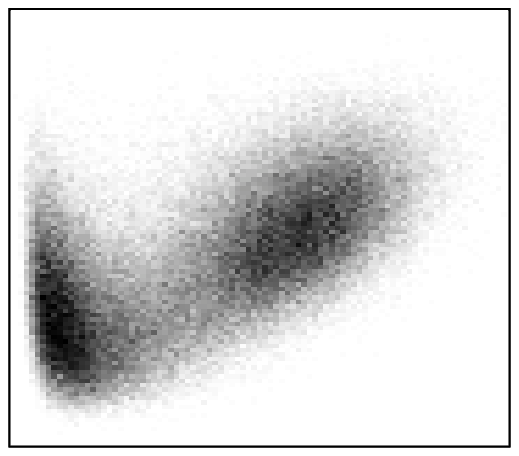}{}
\histbox{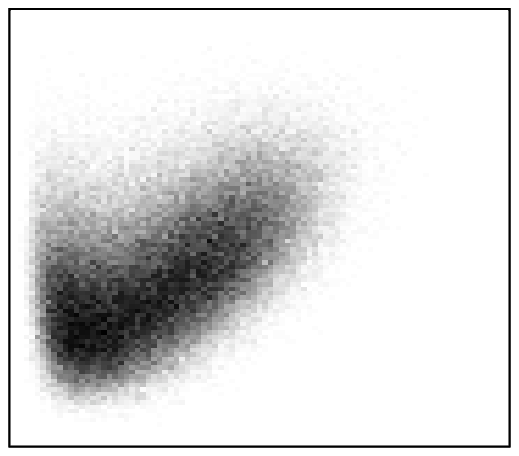}{}
\histbox{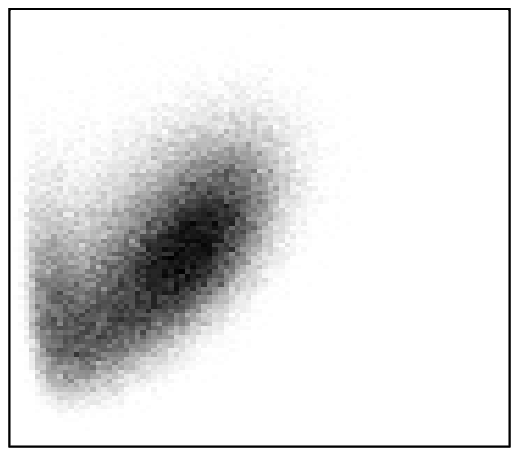}{}
\histbox{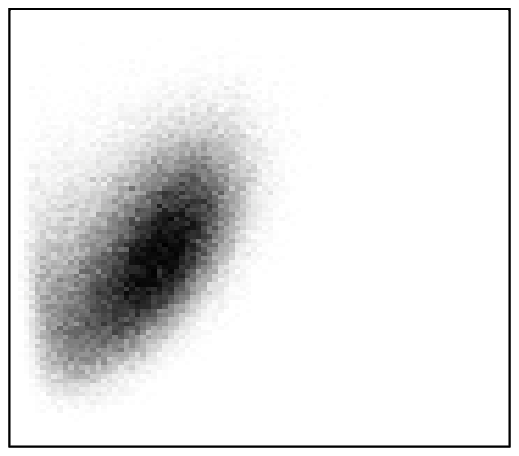}{}
\histbox{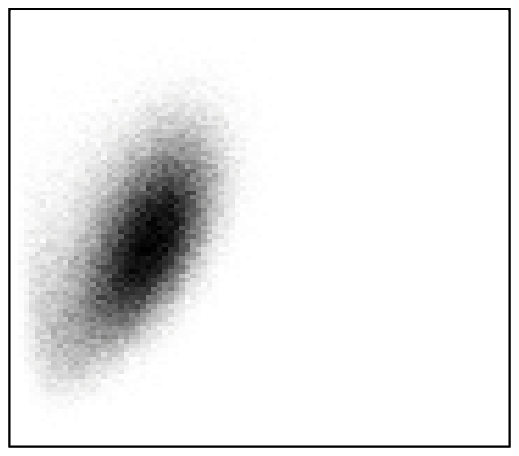}{}
\vspace{.5em}

\noindent
\parbox[b]{.6cm}{\rotatebox{90}{$T=1.05$}}
\histbox{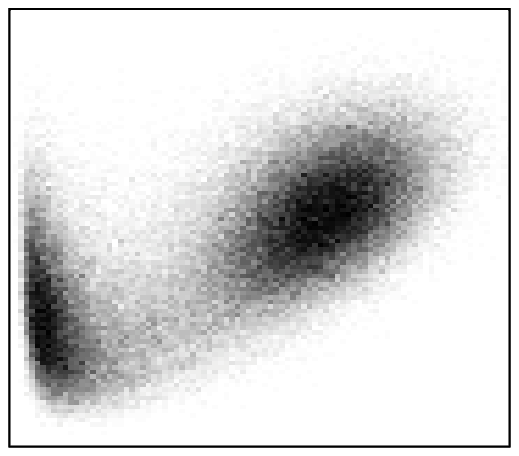}{}
\histbox{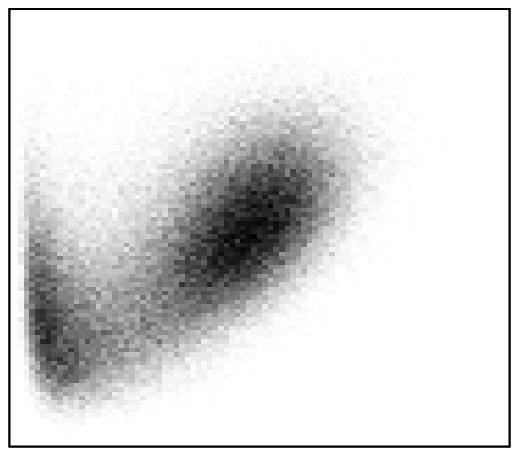}{}
\histbox{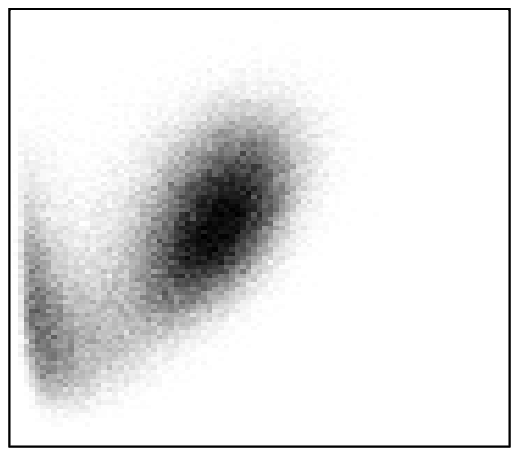}{}
\histbox{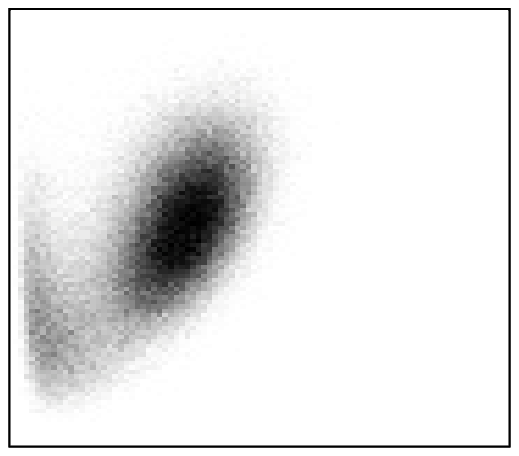}{}
\histbox{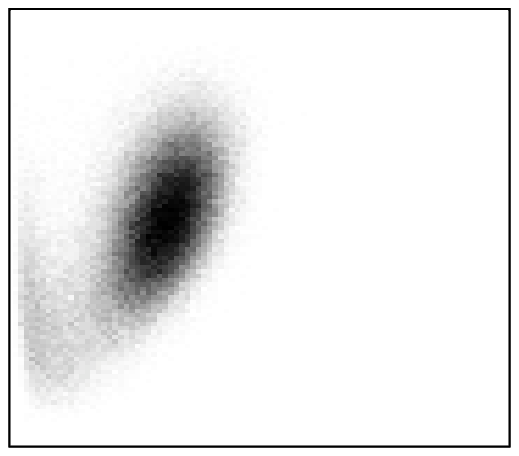}{}
\vspace{.5em}

\noindent
\parbox[b]{.6cm}{\rotatebox{90}{}}
\parbox[b]{.18\linewidth}{\centering $\rho=0.21$}
\parbox[b]{.18\linewidth}{\centering $\rho=0.26$}
\parbox[b]{.18\linewidth}{\centering $\rho=0.31$}
\parbox[b]{.18\linewidth}{\centering $\rho=0.36$}
\parbox[b]{.18\linewidth}{\centering $\rho=0.41$}

\caption{Histograms of $\beta_0^{2,0}$ and $d$ for Lennard-Jones configurations.
The $x$ axis of each box shows local particle number density $d^{(i)}\in[0; 4.5]$ of individual Voronoi cells,
the $y$ axis shows the anisotropy parameter $\beta_0^{2,0}(K^{(i)})\in[0; 1]$.
The critical point would correspond to the configuration at $T=1.1$ and $\rho = 0.26$; below, vapor/liquid coexistence is found;
above, a single fluid phase exists and the probability distribution only has a single component.
}
\label{app:histos-lj-phase-diagram}
\end{figure}
%
%

%
%\begin{figure}
%\centering
%\includegraphics{GlobalAveragePlots/disks_sigma_plot}%
%\includegraphics{GlobalAveragePlots/spheres_sigma_plot}
%
%\caption{Sigma of Beta -- alternative figure}
%\end{figure}
%
%\begin{figure}
%\centering
%\includegraphics{GlobalAveragePlots/disks_sigmaresc_plot}\includegraphics{GlobalAveragePlots/spheres_sigmaresc_plot}
%
%\caption{Sigma of Beta -- alternative figure}
%\end{figure}
%
%

\end{document}